\documentclass{emulateapj} 
\usepackage{graphicx}
\usepackage{color,calc}
\usepackage{amsmath}
\usepackage{amssymb}
\usepackage{amstext}

\newcommand{\Fewbody}{{\em Fewbody\/}}

\begin{document}

\title{Monte Carlo Simulations of Globular Cluster Evolution. IV.\\
  Direct Integration of Strong Interactions}
\shorttitle{Globular Cluster Evolution. IV.}
\submitted{Submitted to ApJ} 
\author{John M. Fregeau\altaffilmark{1} \& Frederic A. Rasio\altaffilmark{2}}
\shortauthors{FREGEAU \& RASIO}
\affil{Department of Physics and Astronomy, Northwestern University, Evanston, IL 60208}
\altaffiltext{1}{fregeau@alum.mit.edu}
\altaffiltext{2}{rasio@northwestern.edu}

\begin{abstract}
We study the dynamical evolution of globular clusters containing populations
of primordial binaries, using our newly updated Monte Carlo cluster evolution
code with the inclusion of direct integration of binary scattering interactions.
We describe the modifications we have made to the code, as well as improvements
we have made to the core Monte Carlo method.  We present several test calculations
to verify the validity of the new code, and perform many comparisons with previous
analytical and numerical work in the literature.  We simulate the evolution of a large
grid of models, with a wide range of initial cluster profiles, and with binary
fractions ranging from 0 to 1, and compare with observations of Galactic
globular clusters.  We find that our code yields very good agreement with
direct $N$-body simulations of clusters with primordial binaries, but yields some 
results that differ significantly from other approximate methods.  Notably, the direct
integration of binary interactions reduces their energy generation rate
relative to the simple recipes used in Paper III, and yields smaller core
radii.  Our
results for the structural parameters of clusters during the binary-burning
phase are now in the tail of the range of parameters for observed clusters, implying that
either clusters are born significantly more or less centrally concentrated than
has been previously considered, or that there are additional physical processes 
beyond two-body relaxation and binary interactions that affect the structural characteristics
of clusters.
\end{abstract}


\section{Introduction}\label{sec:intro}

Observations 
\citep[e.g.,][]{1996AJ....112..574C,2002AJ....123.2541B,1997ApJ...474..701R,2002ASPC..263..163C,2002AJ....123.1509B}, 
in combination with recent theoretical work \citep{2005MNRAS.358..572I},
suggest that although the currently observed binary fractions in the cores of globular clusters 
may be small ($\lesssim 10\%$), the initial cluster binary fraction may have
been significantly larger ($\gtrsim 50\%$).  As has been understood 
theoretically for some time, primordial binaries in star clusters generally
act as an energy source (through super-elastic scattering encounters), 
producing energy in the core and postponing deep core collapse
in a quasi-steady state ``binary-burning'' phase.  
This is analogous to the long-lived main-sequence in stars, in which hydrogen is burned to prevent
collapse.  An initial binary fraction of
a few percent is enough to postpone deep core collapse for many initial relaxation times 
\citep[see][for discussion and references]{2003ApJ...593..772F}.
In addition to playing a large part in the global evolution of a star cluster,
dynamical interactions of binaries also strongly affect the formation and evolution
of stellar and binary exotica, which include low-mass X-ray binaries, recycled pulsars,
cataclysmic variables, and blue stragglers 
\citep{1991A&A...241..137H,1995ApJS...99..609S,1998MNRAS.301...15D,2000ApJ...532L..47R,2005MNRAS.358..572I}.

Similarly to dynamical binary interactions, physical stellar collisions 
(both direct star--star collisions, and those mediated by resonant binary
interactions) also play an important role in the evolution of globular cluster
populations.  Stellar collisions are thought to be one of the two primary
mechanisms by which blue stragglers are created in dense star clusters
\citep[e.g.,][]{2004ApJ...605L..29M}.  A runaway sequence of stellar
collisions of massive main sequence stars early in the lifetime of
a dense cluster may yield a very massive star ($\gtrsim 10^3\,M_\sun$), which
may then become an intermediate mass black hole 
\citep[see, e.g.,][for discussion and references]{2006ApJ...640L..39G}.

The complete picture of star cluster evolution 
can be rather complicated, as it includes, in addition to the two physical processes just mentioned, 
single star evolution, binary star evolution, and tidal stripping due to the field of the Galaxy.  
Even if one ignores these additional processes, modeling
a dense stellar cluster with primordial binaries still presents a formidable
computational challenge.  There are at least two reasons for this: 1) dynamical
interactions of binaries---which are typically resonant, lasting for many
orbits---must be resolved on their natural timescale, which is orders of magnitude shorter
than the cluster relaxation time, and 2) primordial binaries extend the life
of a cluster.  Although the GRAPE series of special-purpose computers is steadily
increasing in performance, direct $N$-body simulation of the 
evolution of clusters with more than a few percent
binaries and a moderate number of stars ($\sim 10^5$) is still quite computationally
expensive, with computational timescales on the order of months.  Faster, more approximate
methods, such as the anisotropic gas model or direct solution of the Fokker-Planck
equation, suffer from the difficulty inherent in incorporating physics beyond
two-body relaxation, such as stellar evolution or binary interactions, in these methods.
The Monte Carlo method bridges the gap between these two computational extremes,
since it allows for the relatively facile inclusion of additional layers of physics, 
provides a star-by-star description of a cluster, and is computationally inexpensive.

In previous studies using approximate methods like Monte Carlo or Fokker-Planck, 
binary interactions had generally been treated using recipes culled
from the results of large numbers of numerical scattering experiments.  (The work of
\citet{2003MNRAS.343..781G}, which incorporates direct integration of binary interactions,
is one notable exception.)  The recipes are typically known only for equal-mass binary interactions, 
thus prohibiting the use of a cluster mass function.  
Thus in order to model realistic clusters, which contain a wide range of masses, one must 
numerically integrate each binary interaction in order to resolve it properly.
We have now incorporated into our Monte Carlo code a dynamical integrator to exactly
integrate dynamical interactions of binaries, allowing us to evolve clusters with
mass spectra, and perform more realistic comparisons with direct $N$-body
calculations.  As we demonstrate below, the direct integration of 
binary interactions reduces their energy generation rate
relative to the simple recipes used in Paper III, and yields smaller core
radii.  

This is the fourth paper in a series studying the evolution of globular clusters
using the Monte Carlo method.  Paper I describes the core method and presents
several test calculations exhibiting the validity of the code \citep{2000ApJ...540..969J}.
Paper II treats the evolution of tidally-truncated clusters with mass spectra
\citep{2001ApJ...550..691J}.  Paper III adds a recipes-based treatment of binary scattering
interactions and considers the evolution of an ensemble of
clusters of varying initial binary fraction and central concentration, finding
that even a small fraction of binaries in a cluster is sufficient to support 
the core against collapse significantly beyond the normal core-collapse time
\citep[][hereafter Paper III]{2003ApJ...593..772F}.  In this, the fourth paper in the series, 
we describe our new code, perform several tests to ensure its validity,
and perform a large set of simulations of clusters with primordial binaries, which we
compare with previous results in the literature and with observations of Galactic globular
clusters.  Section \ref{sec:method} describes our new code in detail, including the additional
physical processes we have added (numerical integration of binary scattering interactions,
and star--star physical collisions), as well as the improvements we have made to the core
Monte Carlo method.  Section \ref{sec:exampleresults} presents a few example results,
and compares them with semi-analytical theory and previous numerical calculations.
Section \ref{sec:results} describes the trends evident in the grid of cluster models
we simulated, and compares our results with observations.  Finally, in section
\ref{sec:summary} we summarize and conclude.

\section{Method}\label{sec:method}

Here we describe in detail our implementation of the Monte Carlo numerical method for simulating 
the evolution of dense star clusters.  It incorporates many physical processes of relevance 
in dense star clusters, including two-body relaxation, direct physical stellar collisions, 
and dynamical interactions of binaries.  Each of these physical processes is treated sufficiently
accurately so as to allow for a rather wide mass spectrum (e.g., $M_{\rm max}/M_{\rm min}\sim 10^3$
for a Salpeter mass function).  For now we neglect stellar evolution (both single and binary), 
but plan to include it in our code in the near future.  Those readers uninterested in the 
technical details of our numerical method can safely skip ahead to the next section.

\subsection{Units}

Before describing our method in detail, we discuss a necessary formality.  We use the standard 
$N$-body system of units \citep{1986LNP...267..233H,2003gmbp.book.....H}. 
For reasons of convenience, we use two units of time in our code.  In addition
to the standard $N$-body unit of time (which is roughly the crossing time), we use
the relaxation time\footnote{Using the relaxation time as the time unit
removes $N$ from any equations of relaxational evolution.}.  
Thus our full (over-specified) system of units is given by the standard 
formulae:
\begin{eqnarray}
  U_m &=& M_0\\
  U_l &=& \frac{G M_0^2}{-4E_0} \\
  U_t &=& \frac{GM_0^{5/2}}{(-4E_0)^{3/2}}\\
  U_{t_{\rm rel}} &=& \frac{N_0}{\ln\gamma N_0} U_t \, , \label{eq:Utrel}
\end{eqnarray}
where $M_0$ is the initial total mass of the cluster, $E_0$ is the initial total
energy of the cluster, $N_0$ is the initial number of stars, and $\gamma N_0$ is 
the Coulomb logarithm.  The quantity $\gamma$ is a function of the initial 
structure of the cluster, and is only needed when converting time in code
units to physical units.  Thus it does not need to be specified for purely
relaxation calculations, which can be quoted in units of the relaxation time, but it
does need to be specified for calculations which include additional physics, like
physical collisions and binary interactions.  For our simulations we
set $\gamma$ via comparisons with $N$-body results (where available), as discussed 
later.

\subsection{Standard Definitions}

It is often useful to be explicit about how certain derived measurable quantities 
are calculated.  For the half-mass relaxation time we adopt the standard definition
\citep{1971ApJ...164..399S}:
\begin{equation}
  t_{\rm rh} = \frac{0.138 N}{\ln(\gamma N)}\left(\frac{r_h^3}{GM}\right)^{1/2} \, ,
\end{equation}
where $N$ is the number of bound cluster objects (single star or binary), $r_h$ is the
radius containing half the mass of the cluster, and $M$ is the total cluster mass.  
Most plots presented in this paper use as time unit the initial half-mass relaxation
time, i.e.\ $t_{\rm rh}$ evaluated at the start of the simulation.  For the core radius
we use the density-weighted average described in \citet{1985ApJ...298...80C}, with
$j=6$ and where the averaging is performed from the cluster center out to the half
mass radius.  In this scheme, a local density is estimated for each star in the cluster
by taking the average density within a sphere centered on the star with radius
at the $j$th star.  The core radius is then estimated by taking the local density weighted
average of star position out to the half mass radius.
We use the same density-weighted averaging scheme to estimate the central
density, velocity dispersion, and average mass, except in places where otherwise noted.
In some cases, as described in the text, we also plot the core radius as measured by the 
standard definition \citep{1987degc.book.....S}:
\begin{equation}
  r_c = \left(\frac{3\sigma_0^2}{4\pi G \rho_0}\right)^{1/2} \, ,
\end{equation}
where $\sigma_0$ is the central three-dimensional velocity dispersion,
and $\rho_0$ is the central mass density.  Note that the results we compare with
below from \citet{2006MNRAS.368..677H} adopt the density-weighted average definition
of the core radius.  As shown in Fig.~\ref{fig:gao.rcrh}, the two definitions differ
minimally when the cluster is in the binary burning phase.

\subsection{Two-Body Relaxation}\label{subsec:relaxation}

Two-body relaxation is the primary physical process responsible for the diffusion
of energy in a star cluster, and thus for its global evolution \citep{2003gmbp.book.....H}.
We use the H\'enon orbit-averaged Monte Carlo method to simulate
two-body relaxation \citep{1971Ap&SS..14..151H}.  For a detailed description
of the basic method we employ, see \citet{2000ApJ...540..969J}.  In summary,
a timestep in the code consists of the following:

\begin{enumerate}
\item Using each star's radial position $r$ and mass $m$, the potential $\Phi(r)$ is calculated
  under the assumption of spherical symmetry (each star is represented by an 
  infinitesimally thin spherical shell).
\item Each pair of stars neighboring in radius undergoes a hyperbolic 
  ``super''-encounter, with scattering angle chosen so as to represent the
  cumulative effect on each star of many long-range, small-angle two-body scattering 
  encounters with all other stars in the system.
\item Using the new radial velocity $v_r$ and tangential velocity $v_t$, the new specific 
  energy $E$ and angular momentum $J$ of each star is calculated (using $\Phi(r)$ from step 1).
\item A new position and corresponding velocity is chosen for each star by picking a point on its orbit randomly,
  sampled in accordance with the amount of time spent at each radial position (i.e., weighted
  by $1/v_r$).
\end{enumerate}

We have made two improvements to the fundamental method which were necessary
to accurately treat star clusters with wide mass spectra, and for the stability
of the long-term evolutions needed for clusters with primordial binaries.

The first improvement is a rather simple one that provides for self-consistency
in step 4 above.  Solving for the new position of a star along its orbit appears
to be a straightforward matter: using the potential calculated in step 1, one writes down
the energy equation for the orbit, $E=\Phi(r)+J^2/2r^2+\frac{1}{2}v_r^2$, 
solves for the pericenter and apocenter, then
samples the radial position with a weighting inversely proportional to 
$v_r$.  However, the potential calculated in step 1 includes the contribution
from the star whose orbit we are trying to solve.  In other words, the star on its orbit
feels the gravitational effect of itself at its old position.  This inconsistency is ignored
in the standard Monte Carlo method \citep{1971Ap&SS..14..151H,2000ApJ...540..969J}, although
it is corrected in the new Monte Carlo code of \citet{2001A&A...375..711F}.  Neglecting this inconsistency
for the case of equal-mass clusters produces a minimal effect on the overall evolution,
slightly postponing core collapse, and leading to a steady drift in the total
system energy.  However, when one considers clusters with even modestly wide mass
spectra ($\sim 0.1$--$10\,M_\sun$), the errors are much less benign.  Since the most massive
stars in the mass spectrum contribute proportionately more to the cluster potential,
it is the calculation of their orbits that is most inaccurate.  In our simulations
we have seen that as the mass spectrum is widened, the core collapse
time gets progressively longer than what would be expected---from direct
$N$-body results and the Monte Carlo calculations of \citet{2006MNRAS.368..121F}---until
it is prevented completely.  In other words, this inconsistency acts as a spurious energy
source, postponing core collapse.  

Correcting the inconsistency in the potential is straightforward.  We simply add a 
correction term to the potential when solving for a star's orbit.  For star
$j$ with mass $m_j$, originally at position $r_j$ when the potential
was last calculated, the correction term is
\begin{equation}
  \Phi_s(r) =
  \begin{cases}
    \displaystyle\frac{Gm_j}{r}& r\geq r_j\\
    \displaystyle\frac{Gm_j}{r_j}& r<r_j\\
  \end{cases}
  \, .
\end{equation}
In principle, for total consistency, one could also add the self-gravity
of the star, $-Gm_j/2r$, since it is treated as a spherical shell.  
However, we find that adding such a term
leads to unphysical behavior for clusters with wide mass spectra when the 
orbit of one of the more massive stars lies within the innermost few
stars (and thus the approximation of spherical symmetry breaks down), the result being that
the star acts as an energy source, ultimately preventing core collapse.  For
narrow mass functions, the addition of the self-gravity term has no noticeable effect 
on the evolution, as found by \citet{2001A&A...375..711F}.  Note that
for the results presented in \citet{2006MNRAS.368..121F} and \citet{2006MNRAS.368..141F}, 
which consider core collapse for wide mass spectra, the self-gravity
term is not included (Freitag 2005, private communication).

The second improvement to the code concerns energy conservation and the long-term
stability of the code.  From the description of a timestep above, it is clear that 
the potential used to find the new positions of stars in their orbits (in step 4) lags behind 
by a timestep.  The result is a steady drift in the 
total system energy.  This can be compensated for by a technique that considers
the mechanical work done by the potential (since it is changing with time) on each
star in the system, and uses it to more accurately calculate the velocities
at the new position on the orbit \citep[see][for details]{1982AcA....32...63S}.  
Briefly, in this method the new specific kinetic energy of each star at its updated position in step 4, 
$\frac{1}{2}v_{\rm new}^2$, is augmented by a term that corresponds to the mechanical work done by the potential
on the star.  The tangential
component of the new velocity is set according to angular momentum conservation as
$v_{\rm t,new} = J / r_{\rm new}$, where $J$ is evaluated in step 3.  This is the same way in which
it is set in the standard Monte Carlo method.
The radial velocity is then simply $v_{\rm r,new} = (v_{\rm new}^2-v_{\rm t,new}^2)^{1/2}$.
Note the asymmetry in which the components of the new velocity are set.  Clearly only the
radial component of the velocity takes into account the mechanical work done by the potential.
Thus it can happen that $v_{\rm new}^2 < v_{\rm t,new}^2$, yielding a nonsensical result
for $v_{\rm r,new}$, in which case ad hoc prescriptions must
be used, sometimes leading to spurious results.  We use a modified
version of this method in which we simply 
preserve the ratio $v_{\rm r,new}/v_{\rm t,new}$ as predicted by the standard Monte Carlo method and scale
the velocities so that $v_{\rm r,new}^2 + v_{\rm t,new}^2 = v_{\rm new}^2$.  This technique appears
to violate angular momentum conservation for individual orbits, but since the Monte Carlo
method assumes spherical symmetry, the total angular momentum of the system
remains statistically consistent with zero.  Our modification to the method of 
\citet{1982AcA....32...63S} yields
results for clusters with mass spectra that are more consistent with direct $N$-body simulations,
and the simulations of \citet{2006MNRAS.368..121F}.  Moreover, the method provides for improved energy
conservation throughout long cluster runs, typically conserving energy to within a part
in $\sim 10^3$ over tens of half-mass relaxation times.

\begin{figure}
  \begin{center}
    \includegraphics[width=\columnwidth]{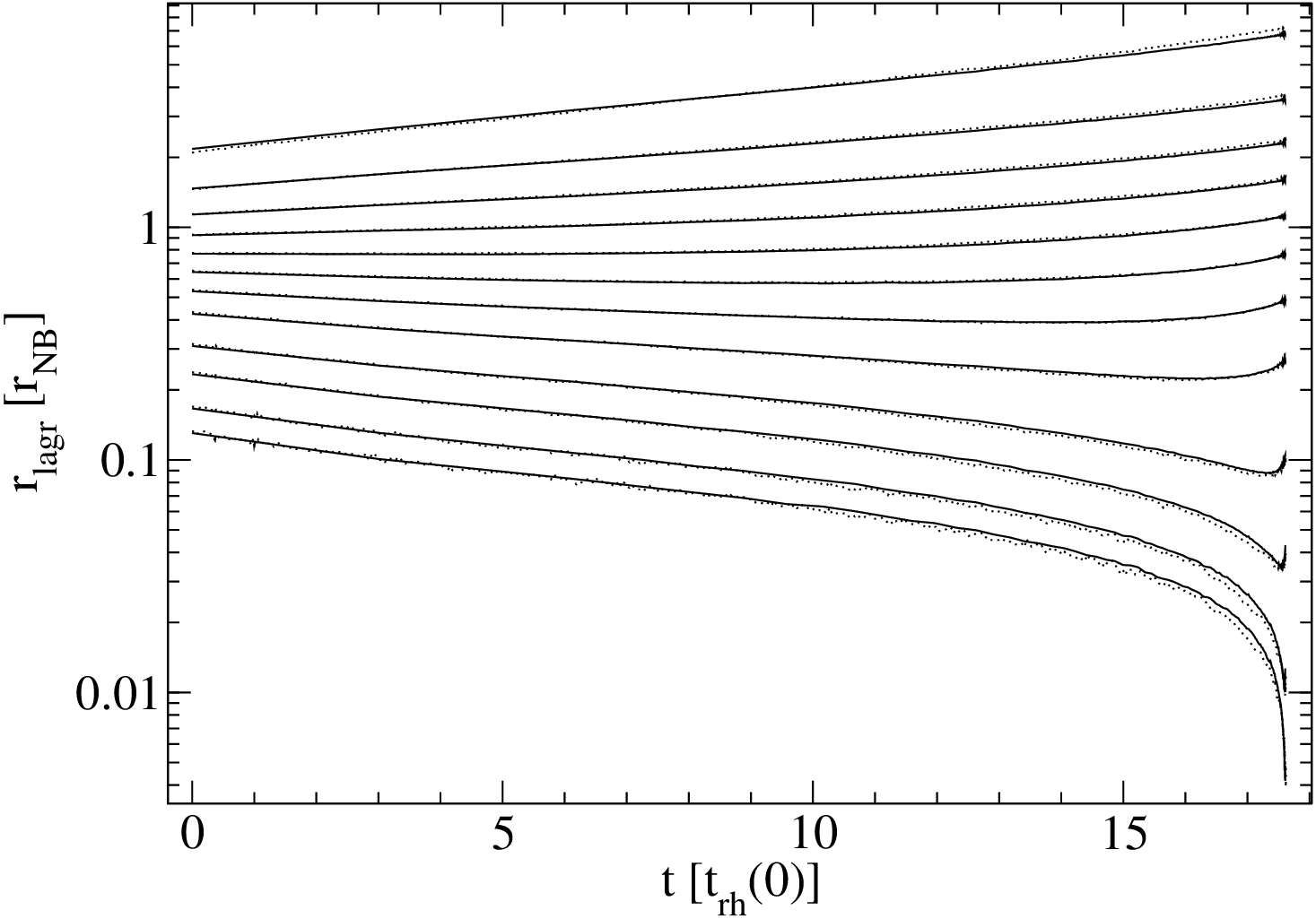}
    \caption{Evolution of the Lagrange radii for a single-component Plummer model (model T1
      in Table~\ref{tab:models}) calculated with
      our Monte Carlo code (solid lines), and compared with a direct $N$-body calculation (dotted lines).
      The Lagrange radii shown enclose a fixed fraction of the total bound
      cluster mass of (from bottom to top) 0.01, 0.02, 0.05, 0.1, 0.2, 0.3, 0.4, 0.5, 0.6, 0.7, 0.8, and 0.9.
      Our model had $5 \times 10^5$ stars, while the $N$-body model had $65536$.  The time unit in the $N$-body
      model was converted from dynamical times to relaxation times using a value of
      $\gamma=0.10$ in the Coulomb logarithm.\label{fig:plummer_lagrad}}
  \end{center}
\end{figure}

To demonstrate the validity of our improved technique for two-body relaxation, we have
compared calculations of clusters subject only to relaxational evolution
with the results from other numerical techniques.  Fig.~\ref{fig:plummer_lagrad}
shows the evolution of the Lagrange radii for a single-component Plummer model (model T1
in Table~\ref{tab:models}) calculated with
our Monte Carlo code (solid lines), and compared with a direct $N$-body code (dotted lines).
The agreement between the Monte Carlo method and direct $N$-body is clearly excellent for this 
model.  The core collapse time of $t_{\rm cc}/t_{\rm rh}=17.6$ is in good agreement with most 
other approximate techniques, as can be seen in the table of \citet{2001A&A...375..711F}.
For the comparison we had to convert the dynamical time units of the $N$-body
model to relaxation time units, using a value of $\gamma=0.10$ in the Coulomb
logarithm.  This value is in good agreement with theoretical arguments
\citep{1975IAUS...69..133H} and other numerical calculations 
\citep{1994MNRAS.268..257G,2001A&A...375..711F}.

\begin{figure}
  \begin{center}
    \includegraphics[width=\columnwidth]{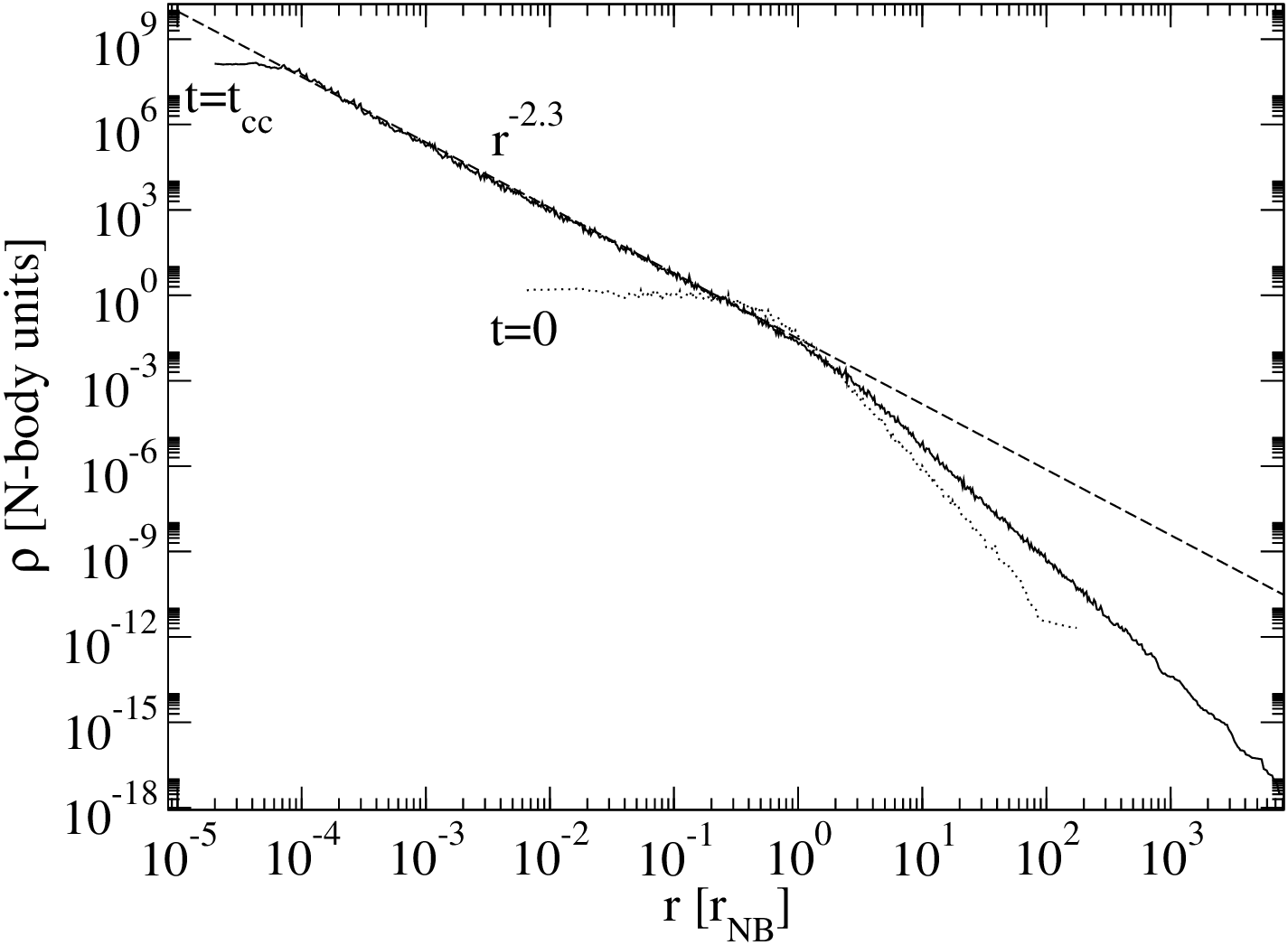}
    \caption{Three-dimensional mass density profiles of model T1 initially and at core collapse.  At core 
      collapse the cluster develops a large power-law profile with exponent $-2.3$, 
      in good agreement with other analytical and numerical calculations.\label{fig:plummer_density}}
  \end{center}
\end{figure}

We have also looked at the evolution of the density profile, as shown in 
Fig.~\ref{fig:plummer_density}, initially and at the time of core collapse.  A power-law 
density profile with $\rho \propto r^{-2.3}$ clearly develops at late times.  The power-law
index of $-2.3$ is in good agreement not only with the results of other Monte Carlo 
calculations \citep{2001A&A...375..711F}, but also with $N$-body simulations which give an index of
$-2.26$ \citep{2003MNRAS.341..247B}, and self-similar analytical Fokker-Planck 
calculations and coarse dynamic renormalization calculations which give
a power-law index of $-2.23$ \citep{2005PhRvL..95h1102S}.

\begin{figure}
  \begin{center}
    \includegraphics[width=\columnwidth]{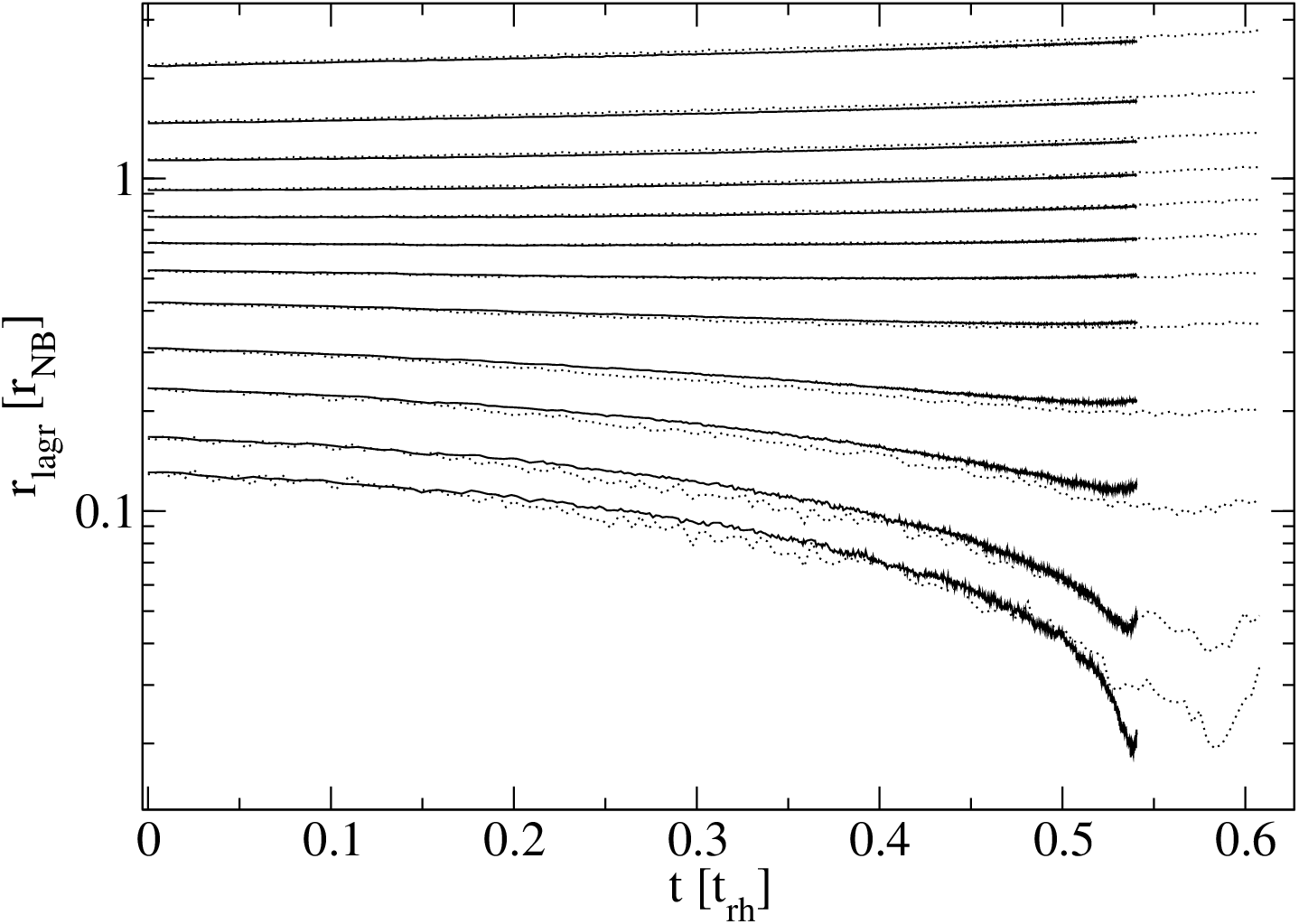}
    \caption{Evolution of the Lagrange radii for a Plummer model with a Kroupa initial mass function
      from $0.1\,M_\sun$ to $10\,M_\sun$ (model T2) calculated with our Monte Carlo code (solid lines), 
      and compared with direct $N$-body (dotted lines).  The Lagrange radii shown enclose a fixed 
      fraction of the total bound cluster mass of (from bottom to top) 0.01, 0.02, 0.05, 0.1, 
      0.2, 0.3, 0.4, 0.5, 0.6, 0.7, 0.8, and 0.9.  Our model had $10^6$ stars, while the $N$-body 
      model had $131072$.  The conversion between dynamical time units and relaxation time units
      required a value of $\gamma=0.05$ in the Coulomb logarithm.\label{fig:plummer_kroupa_lagrad}}
  \end{center}
\end{figure}

\begin{figure}
  \begin{center}
    \includegraphics[width=\columnwidth]{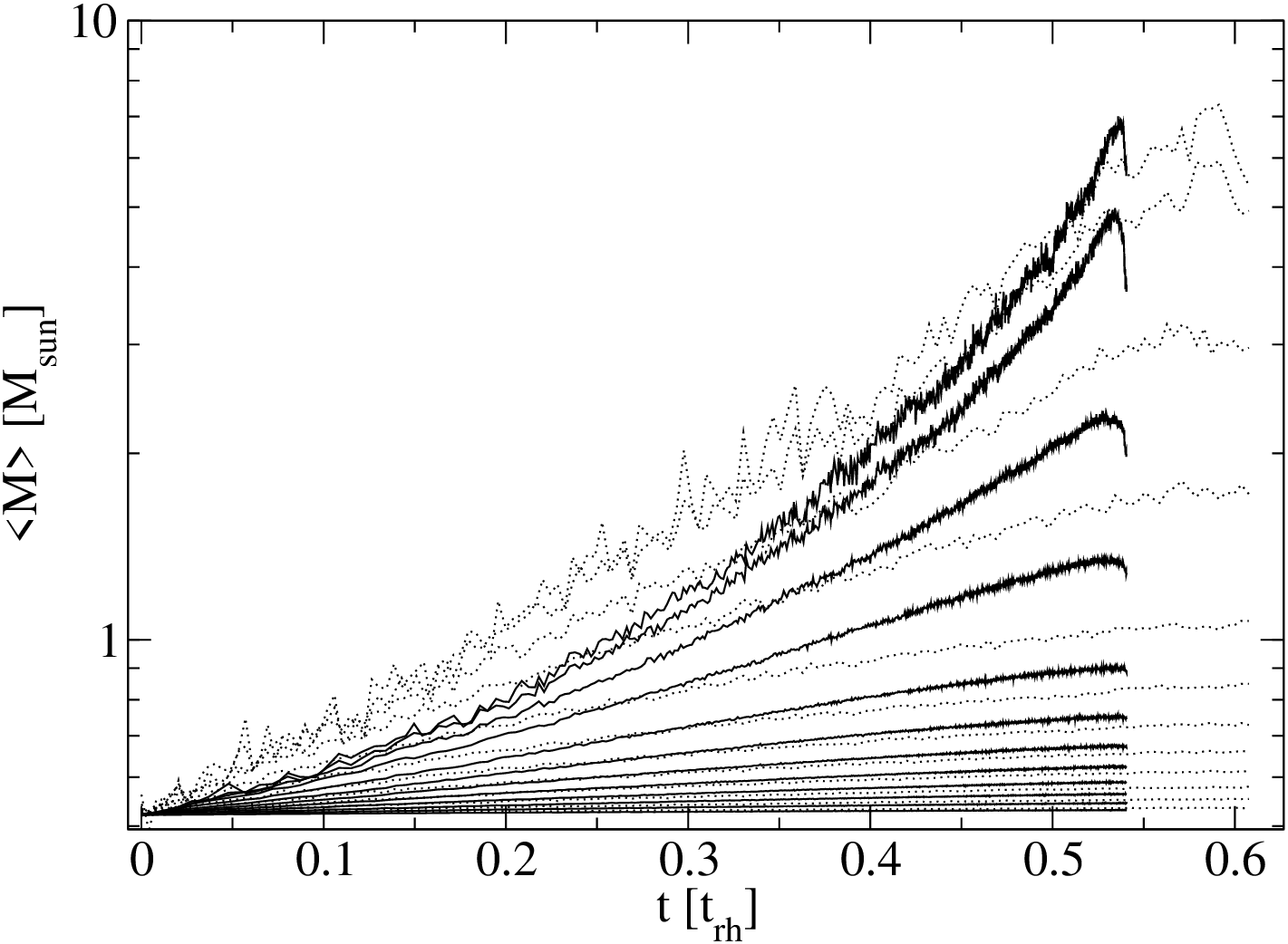}
    \caption{Evolution of the average mass within each Lagrange radius for model T2 
      calculated with our code (solid lines), compared with $N$-body (dotted lines).
      The values of the Lagrange radii are the same as in the previous
      figure.\label{fig:plummer_kroupa_avemass}}
  \end{center}
\end{figure}

Next we considered the evolution of a model with a moderately wide mass spectrum.  
Fig.~\ref{fig:plummer_kroupa_lagrad} compares the evolution of the Lagrange radii calculated 
with our code with a direct $N$-body calculation of a Plummer model with a Kroupa initial 
mass function from $0.1\,M_\sun$ to $10\,M_\sun$ (model T2).  Our model used $N=10^6$ stars while 
the $N$-body model used $N=131072$.  Again, the agreement
is quite good, at least to within the level of noise in the $N$-body simulation.  
The $N$-body model does not undergo deep collapse, since it appears
to form a three-body binary that stalls core collapse.  (Note that
three-body binary formation is not included in our Monte Carlo code.)
A value of $\gamma=0.05$ in the Coulomb logarithm was used to convert between
dynamical time units and relaxation time units.
Fig.~\ref{fig:plummer_kroupa_avemass} shows the evolution of the average mass
within the Lagrange radii, compared with $N$-body.  It appears
that for the innermost Lagrange radii (which have the largest average mass),
our Monte Carlo method predicts an evolution that lags behind the $N$-body method,
but eventually catches up at late times.  Note that the Monte Carlo method of \citet{2006MNRAS.368..121F}
suffers from the same malady, albeit to a lesser degree.

\begin{figure}
  \begin{center}
    \includegraphics[width=\columnwidth]{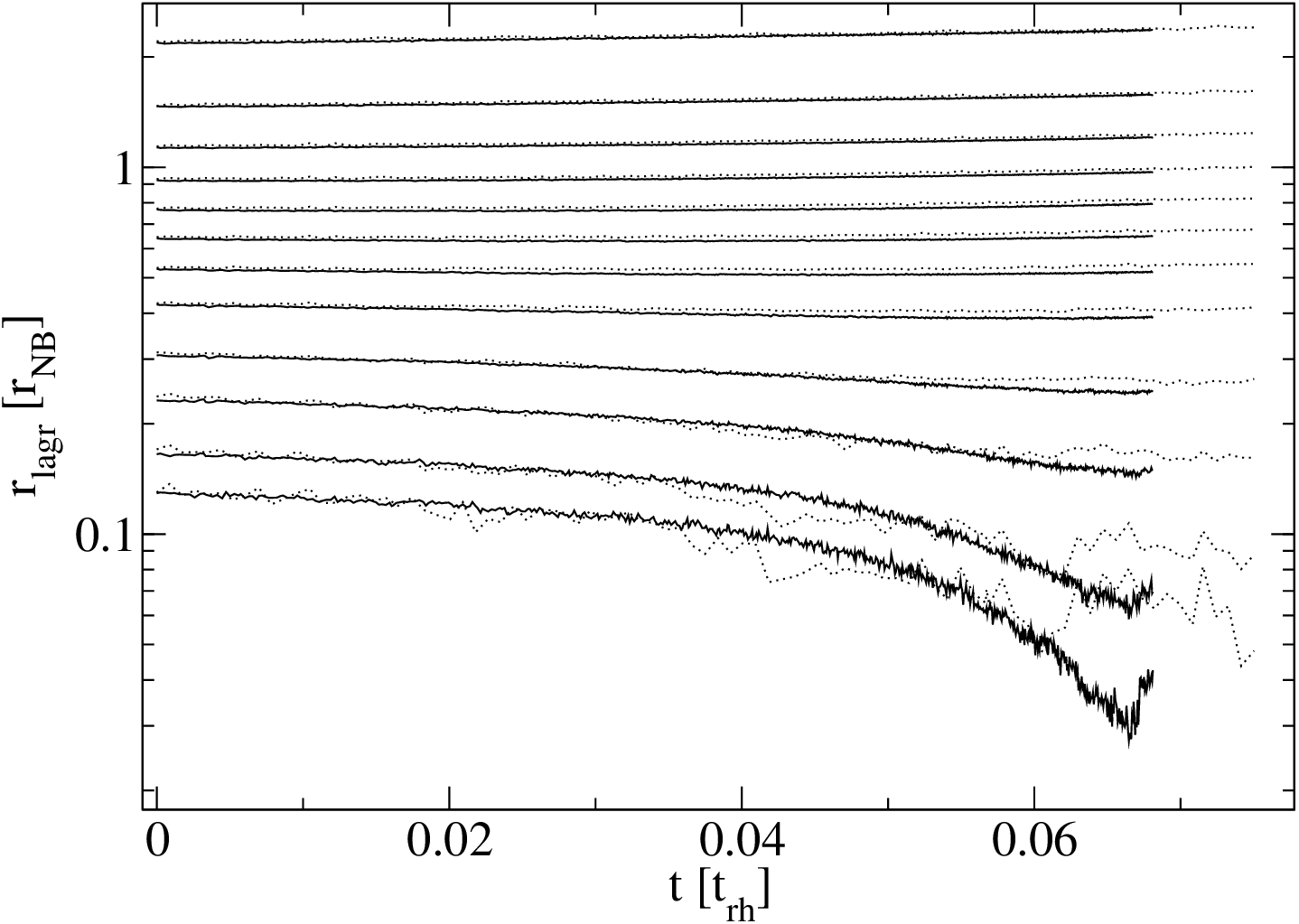}
    \caption{The evolution of the Lagrange radii for a Plummer model with a Salpeter initial 
      mass function from $0.2\,M_\sun$ to $120\,M_\sun$ (model T3) calculated with our Monte Carlo 
      code (solid lines), compared with $N$-body (dotted lines).  The Lagrange radii shown enclose a fixed 
      fraction of the total bound cluster mass of (from bottom to top) 0.01, 0.02, 0.05, 0.1, 
      0.2, 0.3, 0.4, 0.5, 0.6, 0.7, 0.8, and 0.9.  Our model had 
      $1.25\times 10^6$ stars, while the $N$-body model had $262144$.  A value of $\gamma=0.01$ in 
      the Coulomb logarithm was used to convert time units.\label{fig:plummer_salpeter_lagrad}}
  \end{center}
\end{figure}

\begin{figure}
  \begin{center}
    \includegraphics[width=\columnwidth]{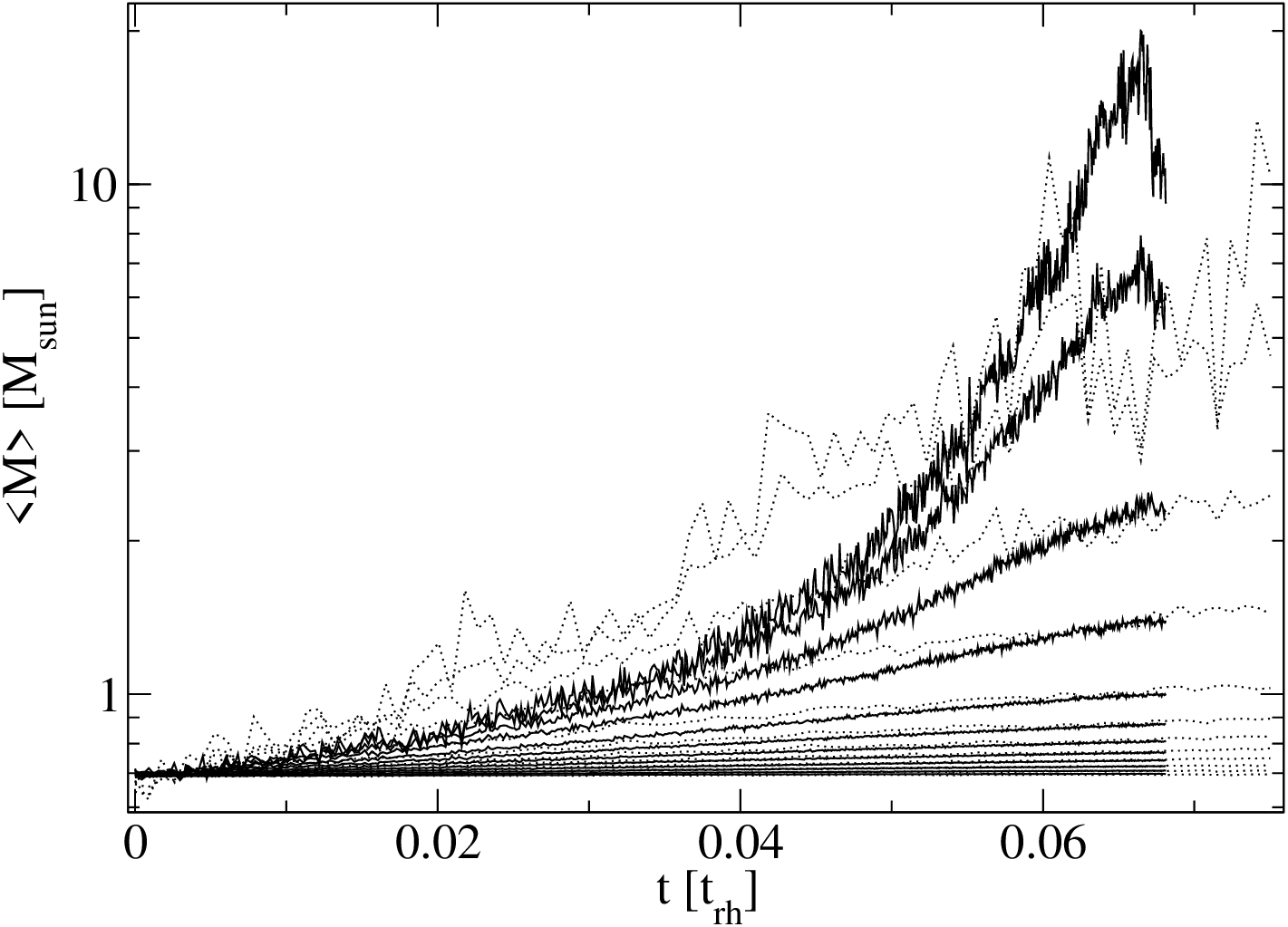}
    \caption{Evolution of the average mass within each Lagrange radius for model T3
      calculated with our code (solid lines), compared with $N$-body (dotted lines).
      The values of the Lagrange radii are the same as in the previous
      figure.\label{fig:plummer_salpeter_avemass}}
  \end{center}
\end{figure}

Finally, we considered the evolution of a model with a very wide mass spectrum.
Fig.~\ref{fig:plummer_salpeter_lagrad} compares the Lagrange radii with $N$-body for a Plummer model
with a Salpeter initial mass function from $0.2\,M_\sun$ to $120\,M_\sun$ (model T3).  Our model
used $N=1.25\times 10^6$ stars while the $N$-body model used $N=262144$.
Again, the agreement is quite good, although the $N$-body model is rather noisy, especially
at late times.  Here a value of $\gamma=0.01$ in the Coulomb logarithm was used,
which is in good agreement with the comparison between Monte Carlo and $N$-body of \citet{2006MNRAS.368..121F}.
Fig.~\ref{fig:plummer_salpeter_avemass} shows the evolution of the average mass within 
the Lagrange radii for the same model.  Here our model lags even further
behind the $N$-body model at early times, but again catches up at late times.  Note that the degree
to which our model disagrees with $N$-body appears to be similar to that of the 
Monte Carlo code of \citet{2006MNRAS.368..121F}.

In general, for two-body relaxation, the agreement of our code with the results of 
direct $N$-body calculations, as well as those of other approximate techniques, has 
been improved greatly by the two major code modifications we have described here.

\subsection{Strong Interactions}\label{subsec:stronginteractions}
Owing to the flexibility of the Monte Carlo method, it is reasonably simple
to layer additional physics on top of the basic two-body relaxation
technique.  This includes those two-body processes which we call 
``strong interactions,'' including single--single physical collisions, dynamical
binary interactions (between two binaries or a binary and a single star), and
large-angle scattering.  At present we have included single--single collisions
and binary interactions.  For both we sample the interactions using the technique
of \citet{2002A&A...394..345F} \citep[also discussed in][]{2001MNRAS.324..218G}.  
In brief, this amounts to evaluating the quantity 
\begin{equation}\label{eq:Pstrong}
  P_{\rm strong}^{(12)}=n_* v_\infty S_{\rm strong}^{(12)} \delta t
\end{equation}
for each pair of stars that neighbor in radial position, where $P_{\rm strong}^{(12)}$ is
the probability for a strong interaction to occur, $n_*$ is the local
number density of stars (binary or single), $v_\infty$ is the relative velocity of the pair at infinity, 
$S_{\rm strong}^{(12)}$ is the cross section for the strong interaction,
$\delta t$ is the timestep, and the notation ``$(12)$'' signifies that the first star
is of type ``1'' while the second is of type ``2''.  
This quantity is a standard ``$n\sigma v$'' estimate for the interaction probability 
between stars of type ``1'' and ``2''.  However, it is the total number density
of stars $n_*$ that appears in the equation, and not $n_1$ or $n_2$.  As shown in 
\citet{2002A&A...394..345F}, when used in this context, eq.~(\ref{eq:Pstrong}) 
yields the correct sampling of the collision rate, since the act of choosing the neighboring star samples the local
density of that type of star \citep[for details, please see section 2.4.2 of][]{2002A&A...394..345F}.

At each timestep, for each pair of stars, the value of $P_{\rm strong}^{(12)}$ is evaluated
($P_{\rm bb}$ for binary--binary interactions, $P_{\rm bs}$ for binary--single,
and $P_{\rm coll}$ for single--single collisions) and compared with a uniform deviate in
$X \in [0,1)$.  If $X < P_{\rm strong}$ the strong interaction is performed, otherwise the pair
undergoes two-body relaxation.

Note that $S_{\rm strong}^{(12)}$ can be written in a very general way in terms of 
the maximum value of the classical pericenter distance between the pair in their 
hyperbolic orbit which yields a strong interaction.  In this case it is
\begin{equation}\label{eq:generalS}
  S^{(12)}_{\rm strong} = \pi b_{\rm max}^2 = \pi r_p^2 \left( 1 + \frac{2GM}{r_p v_\infty^2}\right) \, ,
\end{equation}
where $b_{\rm max}$ is the impact parameter leading to a classical pericenter distance 
of $r_p$, and $M$ is the total mass of the pair.  We will use this expression below.

\subsubsection{Single--Single Collisions}\label{subsubsec:collisions}

Although the simulations we present in this paper do not include physical stellar
collisions (we defer such simulations to a future paper), we still include here
for the sake of completeness a description of the implementation of collisions 
in the code.
For direct physical collisions between main-sequence stars, the outcome can vary 
greatly depending on $v_\infty$.  For relative speeds greater than the escape speed from the 
surface of the star ($v_\infty \ga v_{\rm esc} \approx 500\,{\rm km}/{\rm s}$ for a 
typical solar-mass MS star), which can occur in
galactic nuclei, a collision typically results in a large fraction of the total 
mass lost from the system \citep[see, e.g.,][and references therein]{2005MNRAS.358.1133F}.  
For $v_\infty \la v_{\rm esc}$, which is satisfied for globular clusters, 
the result is typically a clean merger, with a negligible amount of mass lost from the system 
\citep[e.g.,][]{1987ApJ...323..614B,2002ApJ...568..939L}.  Since we are concerned with
the latter case in this paper, we treat physical single--single collisions using the
sticky sphere approximation, which assumes that if the radii of stars touch during
a strong interaction they merge with no mass loss.  The sticky sphere approximation,
when used for stellar collisions of main-sequence stars in models of low velocity dispersion
clusters, has been shown to agree extremely well with the results of more detailed
calculations incorporating the results of SPH simulations \citep{2006MNRAS.368..121F,2006MNRAS.368..141F}.  
With this approximation, the cross section for collisions is given by 
eq.~(\ref{eq:generalS}) with $r_p = R_1 + R_2$:
\begin{equation}
  S^{(12)}_{\rm coll} = \pi (R_1+R_2)^2 \left( 1 + \frac{2GM}{(R_1+R_2) v_\infty^2}\right) \, ,
\end{equation}
where $M$ is the total mass of the two stars.

\begin{figure}
  \begin{center}
    \includegraphics[width=\columnwidth]{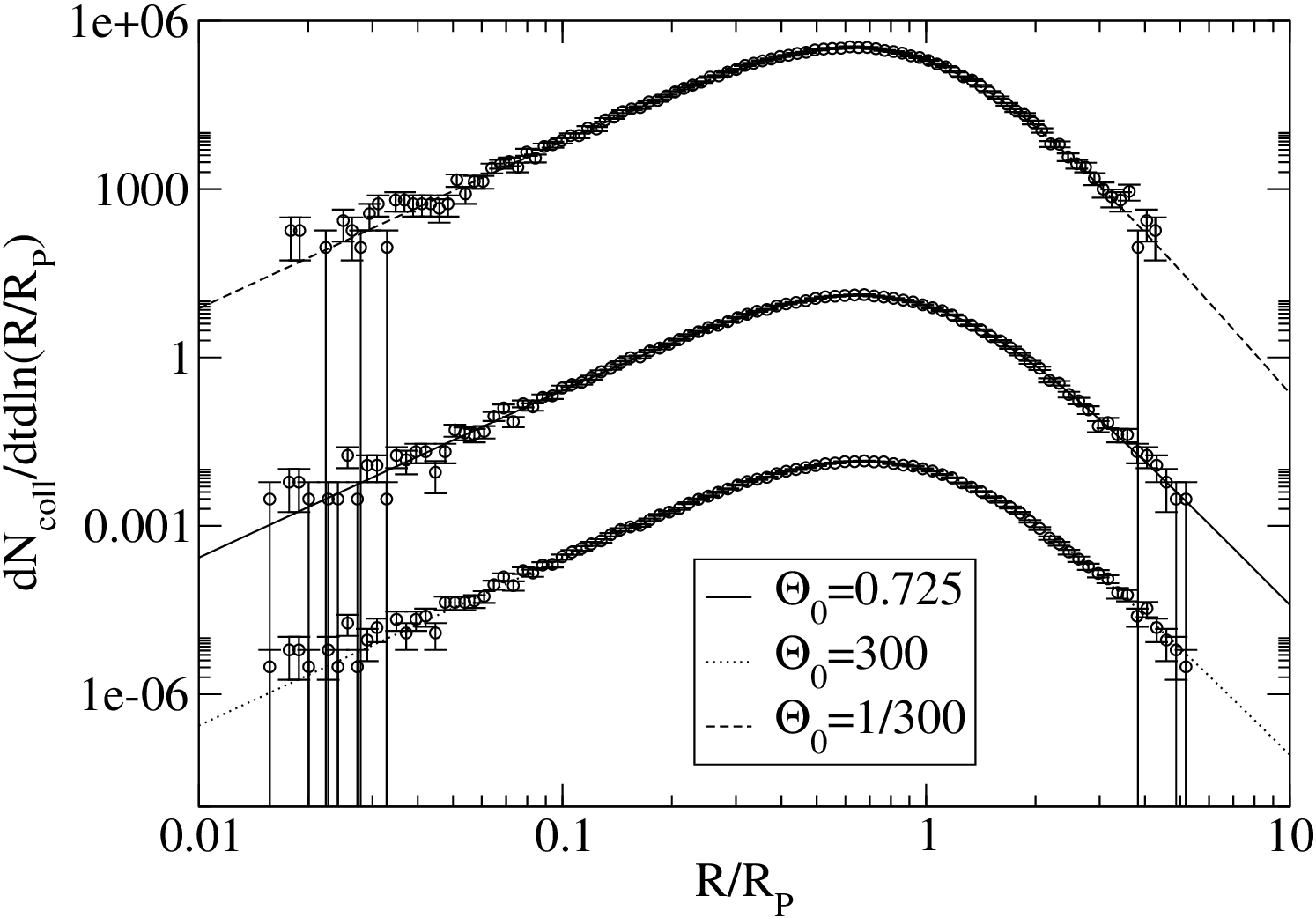}
    \caption{Comparison of the numerically sampled single-single star collision rate (circles) with the 
      analytical result (lines) for a single-mass Plummer model with $N=10^6$ stars, for three different 
      values of the Safronov number ($\Theta_0=1/300$, $0.725$, and $300$).\label{fig:rates}}
  \end{center}
\end{figure}

We have tested that our code correctly samples single--single star collisions
so as to reproduce the correct collision rate.  For a Plummer model, the collision
rate can be solved for analytically, yielding \citep{2002A&A...394..345F}:
\begin{multline}\label{eq:collrate}
  \frac{dN_{\rm coll}(R)}{dt d\ln R} = 54 \left( \frac{3MG}{2\pi R_P^3} \right)^{1/2} u^3 (1+u^2)^{-21/4} \\
  \times\Theta_0^{-2} \left[ 1 + \Theta_0 (1+u^2)^{1/2}\right] \, ,
\end{multline}
where $M$ is the total cluster mass, $R$ is the radial position in the cluster, $R_P$ is the scale
radius of the Plummer model ($R_P=3\pi/16$ in $N$-body units), $u=R/R_P$, and $\Theta_0$ is the 
Safronov number \citep{1987gady.book.....B}.  Based on our sticky sphere collision prescription,
the Safronov number can be written simply as $\Theta_0=3R_P/NR_*$, where $N$ is the number
of stars in the cluster, and $R_*$ is a stellar radius.  We have extracted the collision
rate from a simulation of an unevolving Plummer model (relaxation turned off) composed of 
$N=10^6$ equal-mass stars.  Instead of performing collisions, we simply recorded them.  
Fig.~\ref{fig:rates} shows a comparison of the numerically extracted rate (circles) with the 
analytical rate (lines), for three different values of $\Theta_0$ ($1/300$, $0.725$, and $300$).  
The agreement is excellent for all values of $\Theta_0$.

\subsubsection{Binary Interactions}\label{subsubsec:binints}

All dynamical binary interactions (binary--binary and binary--single) are directly
numerically integrated with \Fewbody, an efficient computational toolkit for evolving small-$N$
dynamical systems \citep{2004MNRAS.352....1F}.  \Fewbody\ was designed specifically
for performing dynamical scattering interactions, and thus it is well-suited for our
purposes.  See \citet{2004MNRAS.352....1F} for a detailed description of the
code.  Note that \Fewbody\ also properly treats physical stellar collisions 
during binary interactions, using the same criterion for a collision as used in 
the Monte Carlo code for single--single collisions.

For sampling binary interactions, we use
the same technique described above for single--single collisions, but with the
cross sections appropriate to binary interactions.  For binary--binary interactions, 
the cross section is given by
eq.~(\ref{eq:generalS}) with $r_p=X_{\rm bb}(a_0+a_1)$, where $a_i$ are the binary
semimajor axes, and $X_{\rm bb}$ is a parameter.  $X_{\rm bb}$ must be set large enough
so that all binary--binary interactions of interest are followed.  Since in this paper we are
concerned mainly with the global evolution of clusters, we need only follow most of
the energy-generating binary interactions.  In principle, one could make $X_{\rm bb}$
arbitrarily large, so as to capture {\em all} potentially interesting interactions.  The 
result would be many more weakly-interacting fly-by interactions, which incur an
infinitesimal computational cost due to \Fewbody's efficient integration techniques.
However, due to the way in which the global Monte Carlo timestep is chosen (see below),
time would grind to a halt in our code.  Clearly, then, setting $X_{\rm bb}$ is a compromise
between capturing all binary interactions of interest (larger $X_{\rm bb}$), and preventing
the timestep from becoming unnaturally small (smaller $X_{\rm bb}$).  We use $X_{\rm bb}=2$
for the results presented in this paper.
For binary--single interactions, we take $r_p=X_{\rm bs}a$, where $a$ is the binary semimajor
axis, and set $X_{\rm bs}=2$.
We find that the values $X_{\rm bs}=X_{\rm bb}=2$ capture almost all the relevant energy-generating
binary interactions.  Test runs with $X_{\rm bs}=X_{\rm bb}=4$ yield values of $r_c/r_h$ in the binary
burning phase that are statistically consistent with $X_{\rm bs}=X_{\rm bb}=2$ runs for both small
and large $f_b$.

\begin{figure}
  \begin{center}
    \includegraphics[width=\columnwidth]{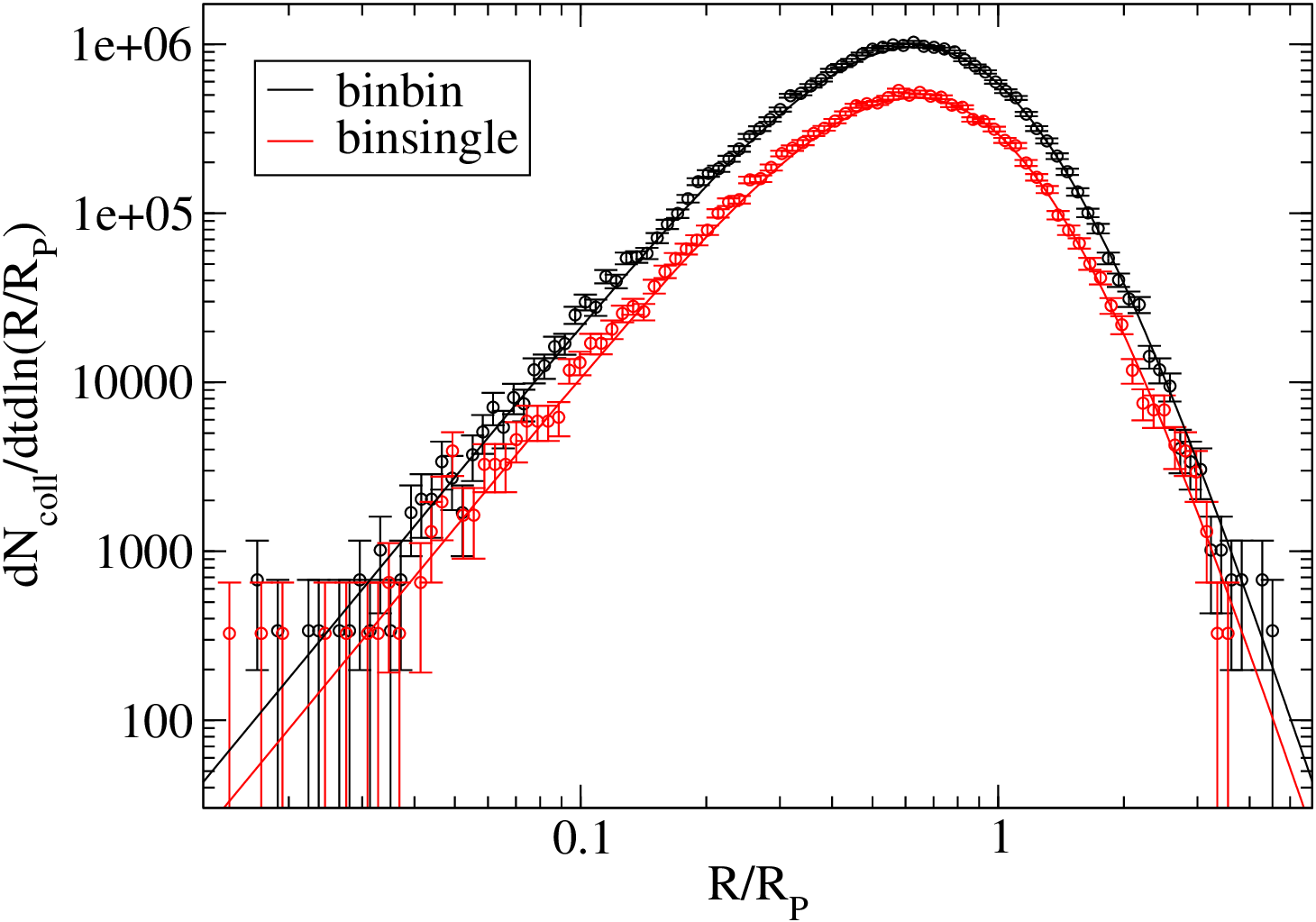}
    \caption{Comparison of the numerically sampled binary interaction rates (circles) with the
      analytical result (lines) for binary--binary interactions (black) and binary--single (red)
      for a single-mass Plummer model with $N=10^6$ stars.  In this test, binaries had the same 
      mass as single stars.  This simplified the analytical calculation of the rate.\label{fig:binrates}}
  \end{center}
\end{figure}

As we did for single--single collisions, we have performed a calculation of an unevolving Plummer model
to test that our code correctly samples the binary interaction rates.  In this test, a fraction
$f_b=0.5$ of the $N=10^6$ stars were binaries.  The binaries had the same mass as single stars
in order to simplify the analytical calculation of the rate.  Again, strong interactions
were not performed, simply recorded.  Since there are two species (binaries and single
stars), the interaction rate is given by eq.~(\ref{eq:collrate}) with an extra factor of
$f_b$ for binary--single, and $f_b^2$ for binary--binary.  Fig.~\ref{fig:binrates}
shows a comparison of the numerically sampled binary interaction rates (circles) with the
analytical result (lines) for binary--binary interactions (black) and binary--single (red).
The agreement is excellent.

Once a binary interaction is deemed to occur, the relative velocity at infinity of the pair is taken
to be the current relative velocity of the pair in the cluster (with the angle of each particle's
tangential velocity randomized), and the impact parameter, $b$,
of the interaction is chosen uniformly in area out to $b_{\rm max}$ as given
in eq.~(\ref{eq:generalS}).  With all parameters of the scattering encounter set,
the interaction is numerically integrated with \Fewbody\ until an unambiguous
outcome is reached.  The outcome products of the interaction are then placed back into the
cluster with their resultant internal properties (mass, binary semimajor axis, eccentricity, etc.)
and external properties (systemic velocity, etc.).  

The only exception to this rule is stable hierarchical triples.  These stable triples 
frequently result from binary--binary interactions 
\citep[roughly 20\% of the time for equal-mass, equal-energy hard binaries; see, e.g.,][]{1983MNRAS.203.1107M}.  
In principle we could keep the triples in the code, and follow their evolution, allowing
them to undergo interactions with other single stars, binaries, and triples (\Fewbody\ can
handle all these cases with ease since it is general in $N$).  However, for simplicity
we currently break triples into a binary and a single star.  We do this by allowing the outer member
of the triple to just barely escape to infinity, with the inner binary shrinking its orbit to conserve
energy in the process.  For simplicity, both the single star and binary are given the systemic velocity
of the original triple.

Finally, we must discuss one more detail.  Since the binary interactions performed with
\Fewbody\ are done in a vacuum---in other words, there are no stars other than the ones
in the interaction to prevent members of the small-$N$ system from making arbitrarily
large excursions---it happens that some binary interactions leave extremely
wide binaries (sometimes as wide as the cluster itself) as their outcome products.
Clearly this is an unphysical situation, as no binary should ever become larger
than the inter-particle separation in a comparable-mass cluster.
We therefore break these pathologically wide binaries at the end of each timestep.
Our criterion for breaking the binaries is that they have an orbital velocity that is roughly
smaller than the local velocity dispersion, since it is this boundary in phase space
and not the hard--soft boundary that determines binary lifetime in a cluster
\citep{2006ApJ...640.1086F}.  We break the binaries if the orbital speed
of their lightest member is less than $X_{\rm hs} \sigma(R)$, where $\sigma(R)$ is the velocity
dispersion at position $R$ in the cluster, and $X_{\rm hs}$ is a parameter which we take to be 0.7.
We set $X_{\rm hs} < 1$ as a safety measure, to ensure that no long-lived binaries
are erroneously broken.

\subsection{Stellar Evolution}\label{subsec:stellarevolution}

For the mass--radius relationship we adopt an approximate piece-wise fit to the
more detailed one used by \citet{2006MNRAS.368..121F} for $Z=10^{-3}$:
\begin{equation}\label{eq:mofr}
    R(M) =
  \begin{cases}
    0.1 R_\sun &\quad M/M_\sun \leq 0.1\\
    R_\sun (M/M_\sun) &\quad 0.1 \leq M/M_\sun \leq 1\\
    R_\sun (M/M_\sun)^{0.57} &\quad 1 \leq M/M_\sun \leq 120\\
    1.6 R_\sun (M/M_\sun)^{0.47} &\quad M/M_\sun \geq 120\\
  \end{cases}
  \, .
\end{equation}
The first two pieces (up to $M = 1\,M_\sun$) are based on \citet{2000ARA&A..38..337C}.  The next
piece is based on \citet{1992A&AS...96..269S}, and the last is from \citet{1984ApJ...280..825B}.
The fit is a rough approximation to the more exact relationships given in the references
listed, but suffices for our purposes since we use it only for determining limits
on the properties of the initial binary population (as described below).

\subsection{Timestep Evaluation}\label{subsec:timestepevaluation}

The timestep in the code should be chosen small enough to resolve the relevant physics 
(two-body relaxation, collisions, binary interactions, etc.), but not smaller than 
necessary.  We take the timestep to be the minimum of the characteristic timescales 
for the different physical processes.

For two-body relaxation, we use the standard expression for the characteristic timescale
for two species of particles undergoing relaxation to deflect each other by an angle
$\theta_{\rm max}$ \citep{2001A&A...375..711F}:
\begin{equation}\label{eq:trel}
  T_{\rm rel} = \frac{\theta_{\rm max}}{\pi/2} \frac{\pi}{32} \frac{v_{\rm rel}^3}{\ln(\gamma N) G^2 n (M_1+M_2)^2} \, ,
\end{equation}
where $v_{\rm rel}$ is the relative speed of the two species, $n$ is the local number density of stars,
and $M_i$ are the masses of each species of star.  The standard expression for the relaxation
time comes from setting $\theta_{\rm max}=\pi/2$ \citep{1987gady.book.....B}.  We 
evaluate eq.~(\ref{eq:trel}) by a local sliding average as
\begin{equation}\label{eq:trelave}
  T_{\rm rel} = \frac{\theta_{\rm max}}{\pi/2} \frac{\pi}{32} \frac{\langle v_{\rm rel}\rangle^3}{\ln(\gamma N) G^2 n \langle (M_1+M_2)^2\rangle} \, ,
\end{equation}
yielding $T_{\rm rel}$ as a function of radial position in the cluster.  We take the minimum value
of $T_{\rm rel}$ for the calculation of the timestep.  The minimum most often occurs at the center 
of the cluster, where the density is the highest.  However, it can sometimes happen for clusters 
with wide mass spectra that the minimum occurs away from the center, due to a massive star in a sea 
of lighter stars.  We adopt $\theta_{\rm max}=1$ for all simulations presented in this paper, which
we find to be a good compromise between accuracy and computational speed.  The distribution of 
scattering angles in a typical timestep has a very long tail at large $\theta$, so most super-encounters 
have a much smaller scattering angle than $\theta_{\rm max}$.

For strong interactions, we evaluate the timescale for each pair of particles neighboring in radius
to undergo a strong interaction, by performing an ``$n\sigma v$'' estimate.  This can be written
\begin{equation}
  T_{\rm strong}^{-1} = \frac{1}{n} \int d^3 {\bf v}_1 d^3 {\bf v}_2 f({\bf v}_1) f({\bf v}_2) |{\bf v}_2-{\bf v}_1| S_{\rm strong} \, ,
\end{equation}
where ${\bf v}_i$ is the velocity of star $i$, and $f$ is the velocity distribution function.
Assuming a Maxwellian velocity distribution for stars ``1'' and ``2'', the result is
\begin{equation}
  T_{\rm strong}^{-1} = 4 \sqrt{\pi} n r_p^2 \sigma \left( 1 + \frac{G M}{2r_p\sigma^2} \right) \, ,
\end{equation}
where $\sigma$ is the one-dimensional velocity dispersion, and we have substituted 
eq.~(\ref{eq:generalS}) for $S_{\rm strong}$.  For collisions, we plug in $S_{\rm coll}$ to find
\begin{equation}
  T_{\rm coll}^{-1} =  16\sqrt{\pi} n_{\rm single} \langle R_*^2 \rangle \sigma 
  \left(1+\frac{G\langle MR_*\rangle}{2\sigma^2 \langle R_*^2\rangle}\right)\, ,
\end{equation}
where we explicitly show which quantities we average, and $n_{\rm single}$ is the number 
density of single stars.  Similarly for binary--binary interactions:
\begin{equation}
  T_{\rm bb}^{-1} =  16\sqrt{\pi} n_{\rm bin} X_{\rm bb}^2 \langle a^2 \rangle \sigma 
  \left(1+\frac{G\langle M a\rangle}{2\sigma^2 X_{\rm bb} \langle a^2\rangle}\right)\, ,
\end{equation}
and binary--single interactions:
\begin{equation}
  T_{\rm bs}^{-1} =  4\sqrt{\pi} n_{\rm single} X_{\rm bs}^2 \langle a^2 \rangle \sigma 
  \left(1+\frac{G\langle M \rangle \langle a \rangle}{\sigma^2 X_{\rm bs} \langle a^2\rangle}\right)\, ,
\end{equation}
where $n_{\rm bin}$ is the number density of binaries, and $a$ is the binary semimajor axis.

\subsection{Initial Conditions}\label{subsec:initcond}

For our initial cluster models we use both isolated Plummer models and tidally-truncated
King models of varying concentration.  Our prescription for tidal mass loss is described
in detail in \citet{2001ApJ...550..691J}.  For models with a mass spectrum or binaries we
assume no primordial mass segregation of the heavier components.  We use a binary
fraction $f_b$ from 0 to 1, with the binary fraction defined as $f_b=N_b/(N_s+N_b)$, where
$N_s$ is the number of single stars in the cluster, $N_b$ is the number of binaries,
and $N=N_s+N_b$.

In assigning the initial properties of the binary population, we start with a cluster of
only single stars.  We create each binary by randomly choosing a cluster star to be the
primary member of the binary, and assigning the secondary mass using a flat distribution
for the binary mass ratio $q$ ($dP/dq \propto 1$), truncated at the low end so that the
mass of the secondary is not lower than the minimum of the initial mass function.  With
the masses of both binary members set, the remaining binary properties are set 
according to one of two different schemes.  The first is the scheme that has traditionally 
been used in numerical modeling of dense stellar systems, in which the binary binding energy 
$E_b$ is distributed uniformly in the logarithm ($dP/dE_b \propto E_b^{-1}$), with 
fixed upper and lower limits.  As shown in Table~\ref{tab:models}, we take as limits on the 
binding energy a few $kT$ on the low end to several hundred $kT$ on the high end, where $kT$ 
is the thermal energy in the cluster core, evaluated as 
$\frac{1}{3} \langle m v^2\rangle \approx \langle m \rangle \sigma_c^2$, where $\sigma_c$
is the one-dimensional velocity dispersion in the core.  With the semimajor axis set
by the binding energy, the eccentricity $e$ is set according to the thermal distribution
($dP/de=2e$).  The second scheme is a slightly modified version of the first in which the 
limits on the binding energy and eccentricity are set in a more physical way.  The binding
energy is still distributed as $dP/dE_b \propto E_b^{-1}$, but with the upper limit set to
the binding energy at a semimajor axis of $5(R_1+R_2)$, where $R_i$ are the stellar radii.
The lower limit on the binding energy is set to that of a binary whose lightest member has
orbital speed $X_{\rm hs} \langle v_{\rm rel} \rangle$, where $X_{\rm hs}$ is as defined
in section \ref{subsubsec:binints}, and $\langle v_{\rm rel} \rangle$ is the locally
averaged relative velocity between objects, taken to be $4\sigma/3\pi$, where $\sigma$
is the local three-dimensional velocity dispersion.  The eccentricity is set according
to the thermal distribution with an upper limit set by the pericenter distance.  
Note that when physical limits on $E_b$ are used, the resulting cluster
simulation is no longer scalable in its length, mass, and time units, since adopting
stellar radii sets the physical scale of the system.

\section{Example Results and Comparisons}\label{sec:exampleresults}

Having verified that our code properly treats two-body relaxation and correctly 
samples the interaction rates for strong interactions, we now use it study 
the evolution of more realistic clusters.  We first consider clusters 
of equal-mass stars with primordial binary populations.  Note again that
for all simulations presented in this paper physical stellar collisions
were turned off.  The focus of
this section is on presenting a few illustrative results in
detail, and comparing the results of our newly modified code with those of
other codes, as well as the previous version of our code.

\begin{figure}
  \begin{center}
    \includegraphics[width=\columnwidth]{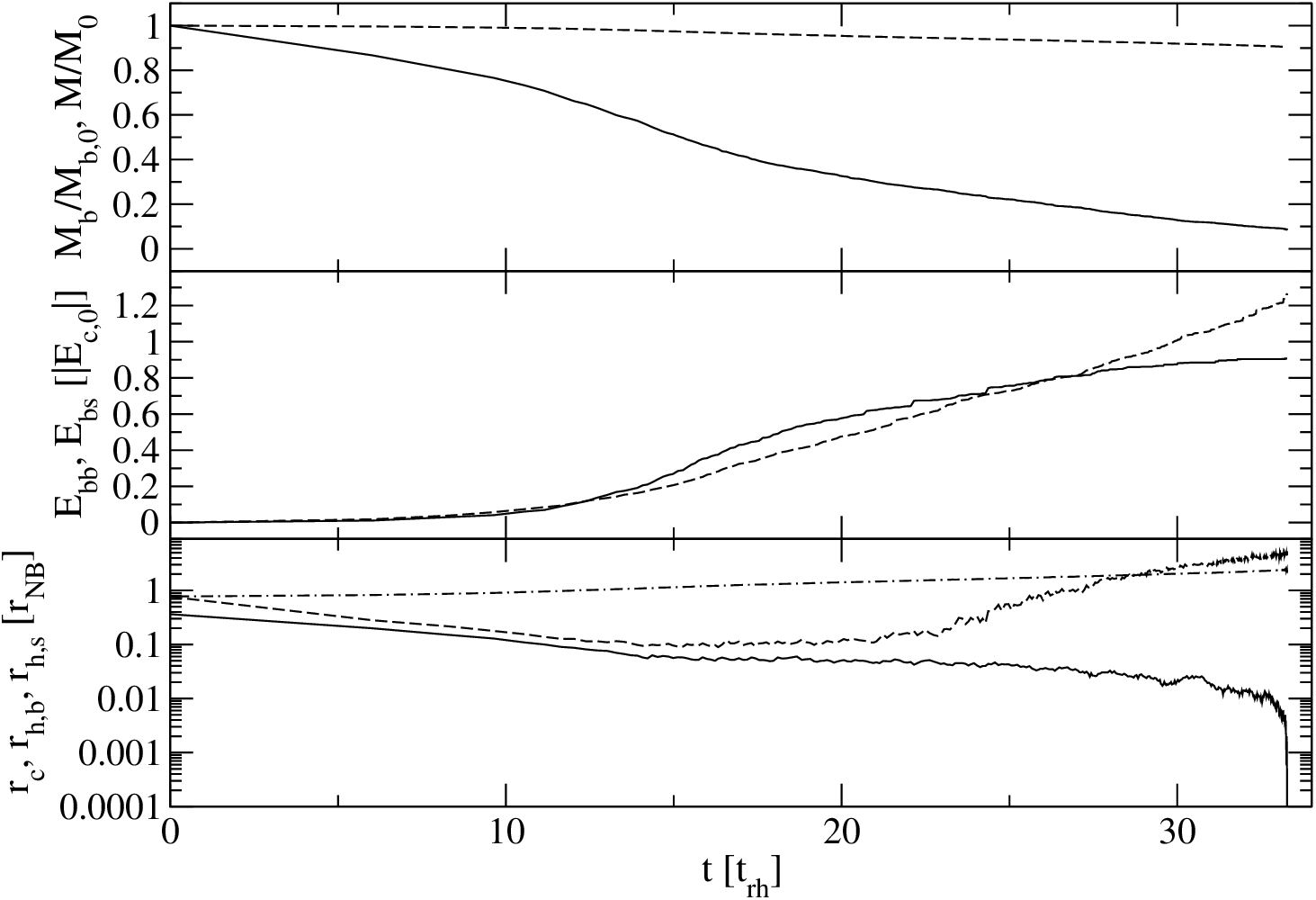}
    \caption{Evolution of an isolated Plummer model with $N=10^5$ stars, and an initial 3\%
      binary fraction (model pl\_n1e5\_fb0.03 in Table~\ref{tab:models}).  The top panel shows $M_b$, the total mass 
      in binaries bound to the cluster (solid line), and $M$, the total mass of the cluster 
      (dashed line) as a function of time, relative to their initial values.  The middle panel 
      shows $E_{\rm bb}$, the cumulative energy generated in binary--binary interactions (solid line) and 
      $E_{\rm bs}$, the cumulative energy generated in binary--single interactions (dashed line) 
      relative to $|E_{c,0}|$, the absolute value of the cluster's initial mechanical energy.  
      The bottom panel shows the evolution of $r_c$, the cluster core radius (solid line), 
      $r_{\rm h,b}$, the half-mass radius of the binaries (dashed line), and $r_{\rm h,s}$, 
      the half-mass radius of single stars (dot-dashed line).
      Time is in initial half-mass relaxation times.\label{fig:pl_n1e5_fb0.03.binary}}
  \end{center}
\end{figure}

Fig.~\ref{fig:pl_n1e5_fb0.03.binary} shows the evolution of an isolated 
Plummer model with $N=10^5$ stars, and an initial 3\%
binary fraction (model pl\_n1e5\_fb0.03 in Table~\ref{tab:models}).  The top panel shows $M_b$, the total mass 
in binaries bound to the cluster (solid line), and $M$, the total mass of the cluster 
(dashed line) as a function of time, relative to their initial values.  The middle panel 
shows $E_{\rm bb}$, the cumulative energy generated in binary--binary interactions (solid line) and 
$E_{\rm bs}$, the cumulative energy generated in binary--single interactions (dashed line) 
relative to $|E_{c,0}|$, the absolute value of the cluster's initial mechanical energy.  The bottom panel 
shows the evolution of $r_c$, the cluster core radius (solid line), $r_{\rm h,b}$, the half-mass radius of
the binaries (dashed line), and $r_{\rm h,s}$, the half-mass radius of single stars (dot-dashed line).
The evolution of this model is typical of a cluster with primordial binaries.  The core
initially shrinks until the central density increases to the point at which energy 
generation in binary interactions is sufficient to prevent the core from collapsing.
The binaries steadily gain binding energy in the subsequent, long-lived binary burning 
phase.  They thus suffer progressively larger kinetic recoil from binary interactions, with the result 
that eventually the half-mass radius of binaries overtakes the single star half-mass radius.  
Eventually the binary population is sufficiently depleted in the core that the core collapses, by
which point a large fraction of the initial binary population
has been depleted ($\sim 90\%$).  The binaries are lost either by being
disrupted or ejected in binary interactions.
From the middle panel it is clear that a significant fraction of energy can be released
via binary interactions, in this case of order the initial cluster mechanical energy.

\begin{figure}
  \begin{center}
    \includegraphics[width=\columnwidth]{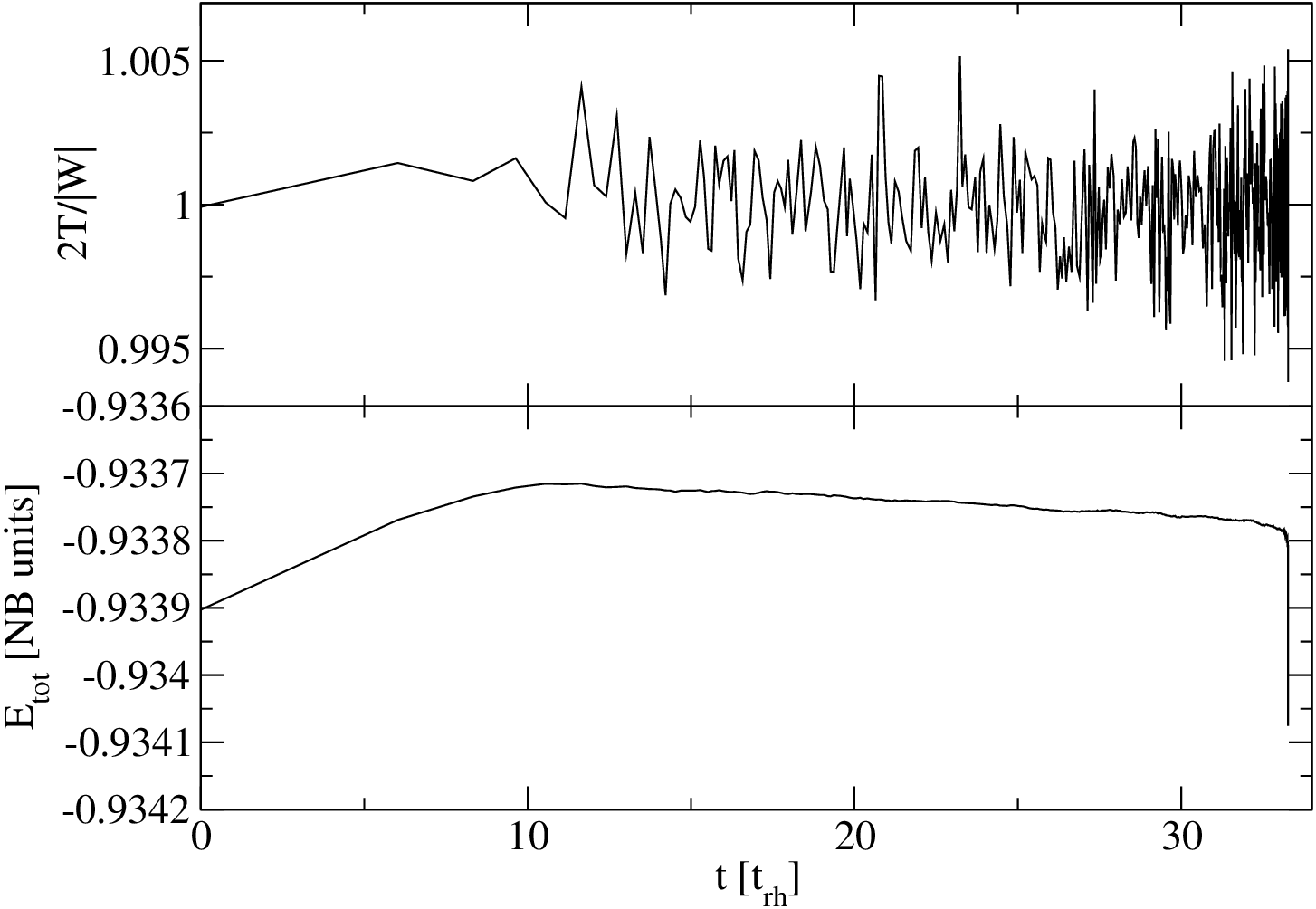}
    \caption{Evolution of the virial ratio (top panel), and total cluster energy (bottom panel) for 
      model pl\_n1e5\_fb0.03.  The energy plotted here includes not
      only the mechanical energy of the cluster, but also the binding energy of the binaries.  Note
      the range on the $y$ axis for the energy plot.  In this run energy was conserved to a part in 
      $\sim 10^4$.\label{fig:pl_n1e5_fb0.03.energy}}
  \end{center}
\end{figure}

Fig.~\ref{fig:pl_n1e5_fb0.03.energy} shows the evolution of the virial ratio and the total
cluster energy (cluster mechanical energy plus binary binding energy) 
for the same model.  The virial ratio is conserved to within statistical fluctuations
for the duration of the calculation, suggesting that the code is yielding
accurate results.  The total cluster energy is also conserved relatively well throughout
the calculation, at the level of a part in $\sim 10^4$, with the largest jump in the energy occurring
during the deep core collapse phase.  The high degree of conservation of energy and of the
virial ratio for this model is typical of all models we present in this paper.

\subsection{Comparison with Theory}\label{sec:theorycomp}

For the first quantitative test of our results for the quasi steady-state binary
burning phase, we appeal to the semi-analytical model of \citet{1994ApJ...431..231V}.
Their model combines the results of binary scattering experiments with
an analytical prescription for energy balance in the core, yielding a relationship
among the core binary fraction, the properties of the binary population, and the
ratio $r_c/r_h$:
\begin{equation}\label{eq:rcrh}
  \frac{r_c}{r_h} = \frac{0.1872}{\log_{10}(\gamma N)}
  \frac{\mu_{\rm bs}\phi_b(1-\phi_b)+\mu_{\rm bb}\phi_b^2}{(1+\phi_b)^4}
  \left(\frac{v_c}{v_h}\right)^3 \left(\frac{\Gamma}{10}\right) \, ,
\end{equation}
where $\phi_b$ is the core binary fraction, $\mu_{\rm bs}$ and $\mu_{\rm bb}$ are
coefficients representing the energy generation rates in binary--single and binary--binary
interactions for a given set of binary properties, $v_c$ and $v_h$ are 
the core and half-mass velocity dispersions, and $\Gamma$ parameterizes the 
expansion rate of the core in terms of the half-mass relaxation time.
\citet{2006MNRAS.368..677H} have shown that
eq.~(\ref{eq:rcrh}) agrees well with the results of $N$-body simulations of $N=4096$
clusters with primordial binaries, with the values $\Gamma=9.4$ and 
$v_c/v_h=\sqrt{2}$, which are close to the canonical values of 
$\Gamma=11.5$ and $v_c/v_h=\sqrt{2}$ adopted in \citet{1994ApJ...431..231V}.
However, they find that the dependence of $r_c/r_h$ on $N$ is steeper than 
eq.~(\ref{eq:rcrh}), in the sense that clusters with $N \gtrsim 10^3$ have a 
systematically smaller value of $r_c/r_h$ than eq.~(\ref{eq:rcrh}) predicts, 
with the opposite true for clusters with fewer stars.  Extrapolating from the 
numerical results in Fig.~18 of \citet{2006MNRAS.368..677H}, it appears that 
for $N \sim 10^5$ eq.~(\ref{eq:rcrh}) overestimates $r_c/r_h$ by a factor
of $\approx 2$ for the values $\Gamma=10$ and $v_c/v_h=\sqrt{2}$.

\begin{figure}
  \begin{center}
    \includegraphics[width=\columnwidth]{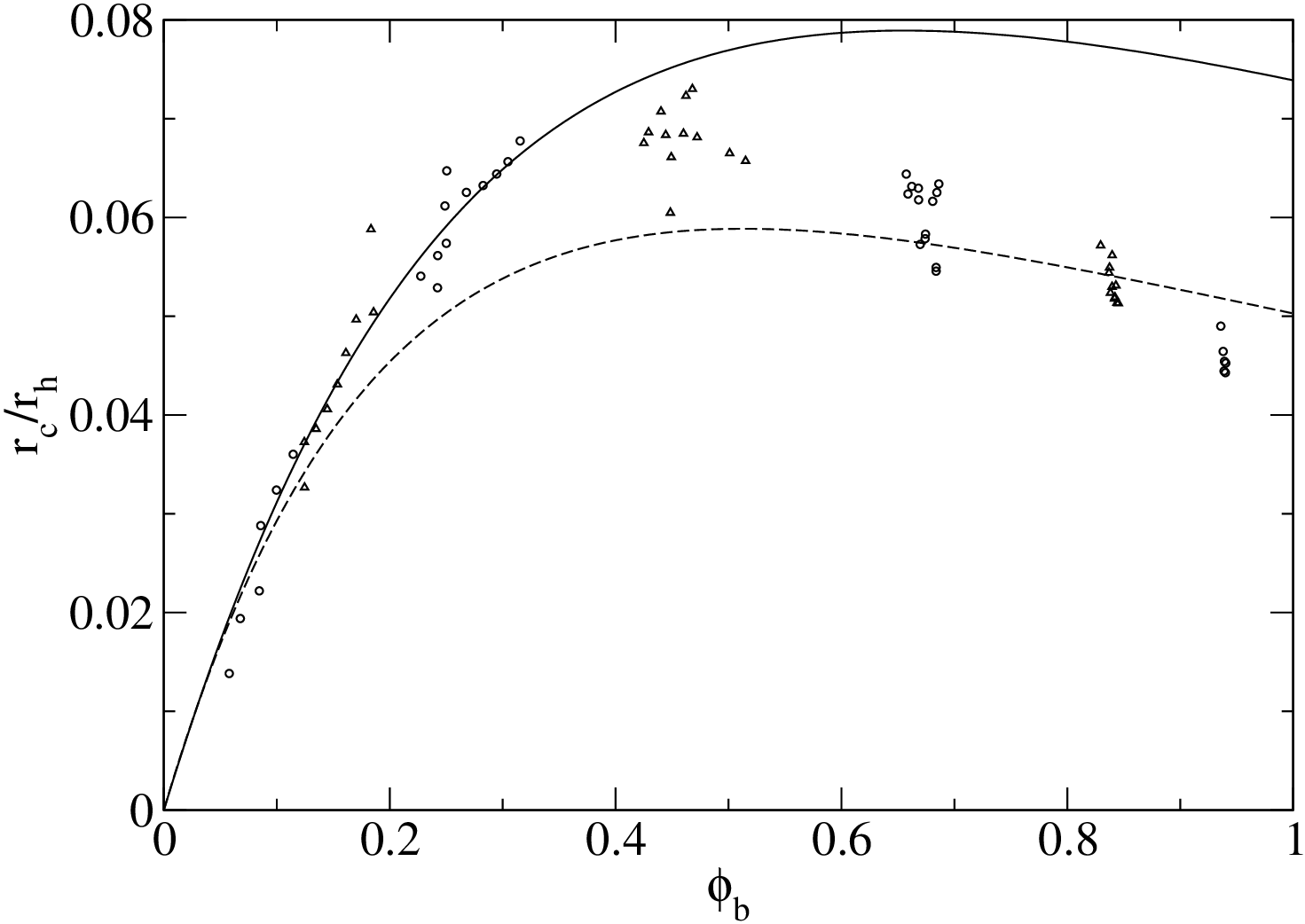}
    \caption{The quantity $r_c/r_h$ vs.~core binary fraction $\phi_b$ for simulations pl\_n3e5\_fb0.01\_kt through
      pl\_n3e5\_fb0.60\_kt, compared with the semi-analytical model of \citet{1994ApJ...431..231V}. 
      Each set of points (shown as circles and triangles, alternating) represents
      a simulation with a different initial binary fraction
      (from left to right): $f_b=0.01$, $0.02$, $0.04$, $0.08$,
      $0.15$, $0.3$, and $0.6$.  Each point is determined by 
      averaging over a time window of width $\Delta t=t_{\rm rh}$ from the point of
      core stabilization until several $t_{\rm rh}$ later.
      The solid line shows the theoretical model with the standard values
      $\Gamma=10$ and $v_c/v_h=\sqrt{2}$ for $E_b=10$--$100\,kT$, while the dashed
      line shows the theoretical model with $E_b=25$--$100\,kT$ to reflect the depletion
      of the widest binaries due to binary--binary interactions.
      \label{fig:vesp_chern}}
  \end{center}
\end{figure}

To test the agreement between our code and eq.~(\ref{eq:rcrh}), we 
have performed several cluster simulations
of varying initial binary fraction and measured the core radius and binary
fraction after core stabilization.  The details of the models (pl\_n3e5\_fb0.01\_kt 
through pl\_n3e5\_fb0.60\_kt) are given in Table~\ref{tab:models}.
Fig.~\ref{fig:vesp_chern} shows $r_c/r_h$ vs.\ core binary fraction $\phi_b$ for
the simulations, compared with eq.~(\ref{eq:rcrh}).  
Each set of points (shown as circles and triangles, alternating) represents
a simulation with a different initial binary fraction.
Each point is determined by averaging over a time window of width 
$\Delta t=t_{\rm rh}$ from the point of core stabilization until several 
$t_{\rm rh}$ later.  The solid line shows the theoretical model with the standard values
$\Gamma=10$ and $v_c/v_h=\sqrt{2}$ for $E_b=10$--$100\,kT$.  The agreement 
between our code and the semi-analytical model is quite satisfactory for 
$\phi_b \lesssim 0.5$.  Above this value the two differ by up to a factor of 
$\sim 2$, with our code yielding $r_c/r_h$ values smaller than predicted by 
theory.  We believe the apparent discrepancy is due to the destruction in 
binary--binary interactions of the wider binaries in the initial binary 
distribution, the rate of which increases
with increasing binary fraction.  For reference, the theoretical curve for
a narrower range of binary binding energies ($E_b=25$--$100\,kT$) is shown
in the dashed line.

Finally, we mention the issue of post-collapse, gravothermal core oscillations.  
Gravothermal oscillations are driven by the gravothermal instability
(essentially the negative heat capacity of gravitational systems, which causes
heat to flow from cold to hot in a runaway fashion).  
At any given time during deep core collapse, the temperature profile is a 
monotonically decreasing function of $r$, with the central temperature steadily
increasing as a function of time.  At some point during the collapse, a binary
scattering interaction occurs, producing a small amount of energy in the core
and cooling it, creating a temperature inversion.  Once the temperature inversion
is established, the gravothermal instability takes over and drives the subsequent
core expansion.  There are several hallmarks of gravothermal oscillations.  
One is, of course, a temperature inversion in the cluster core at the point when 
the core begins to rebound.  Another is that the expansion phase is not driven 
by energy generation in binaries.  Yet another is the presence of loops in a phase
space diagram of the cluster core properties \citep{1996ApJ...471..796M,2006MNRAS.368..677H}.
In Paper III we demonstrated the gravothermal nature of the
core oscillations produced by our code by displaying the temperature inversion
in the core.  As in Paper III, cluster models evolved with our newly modified code
undergo core oscillations after core collapse.  Preliminary tests show
loops in phase space, but with the wrong directional sense, implying that
the oscillations we now see are not gravothermal in nature.  
Since we are concerned only with pre-deep core collapse evolution in
this paper, we postpone for future work a more detailed analysis of the core oscillations
produced by our newly modified code.

\subsection{Comparison with Previous Numerical Work}

The amount of numerical work reported on clusters with primordial binaries has grown
considerably since the work of \citet{1991ApJ...370..567G}, who
used a multimass Fokker-Planck code coupled with a recipes-based treatment
of binary interactions.  In Paper III, we
used a Monte Carlo code coupled with recipes for binary interactions. 
\citet{2003MNRAS.343..781G} used a hybrid code, which treated single
stars via a gas dynamical method and binaries via Monte Carlo,
and performed direct numerical integration of binary interactions.
\citet{2006MNRAS.368..677H} and \citet{astro-ph/0602409} performed
direct $N$-body simulations.  And as described in the preceding sections, in this
paper we use a Monte Carlo method coupled with direct numerical integration
of binary interactions.  In this section we compare the results from our code 
with those from all the methods just listed.

\begin{figure}
  \begin{center}
    \includegraphics[width=\columnwidth]{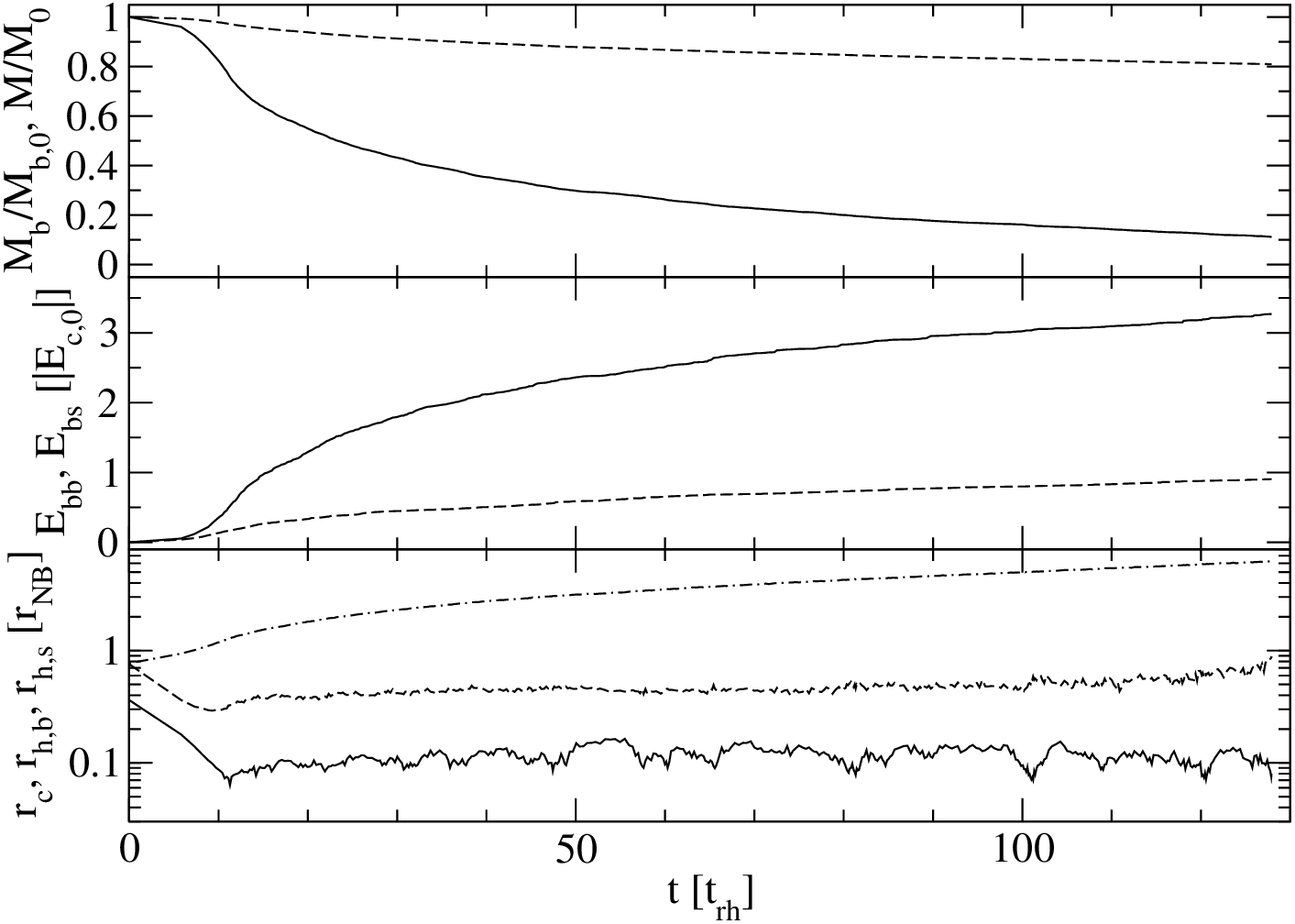}
    \caption{Evolution of the ``Gao, et al.''\ model: an isolated Plummer model with 
      $N=10^5$ objects, equal-mass stars, and 10\% binaries (model T4).
      Quantities plotted are the same as in Fig.~\ref{fig:pl_n1e5_fb0.03.binary}.
      \label{fig:gao.binary}}
  \end{center}
\end{figure}

\begin{figure}
  \begin{center}
    \includegraphics[width=\columnwidth]{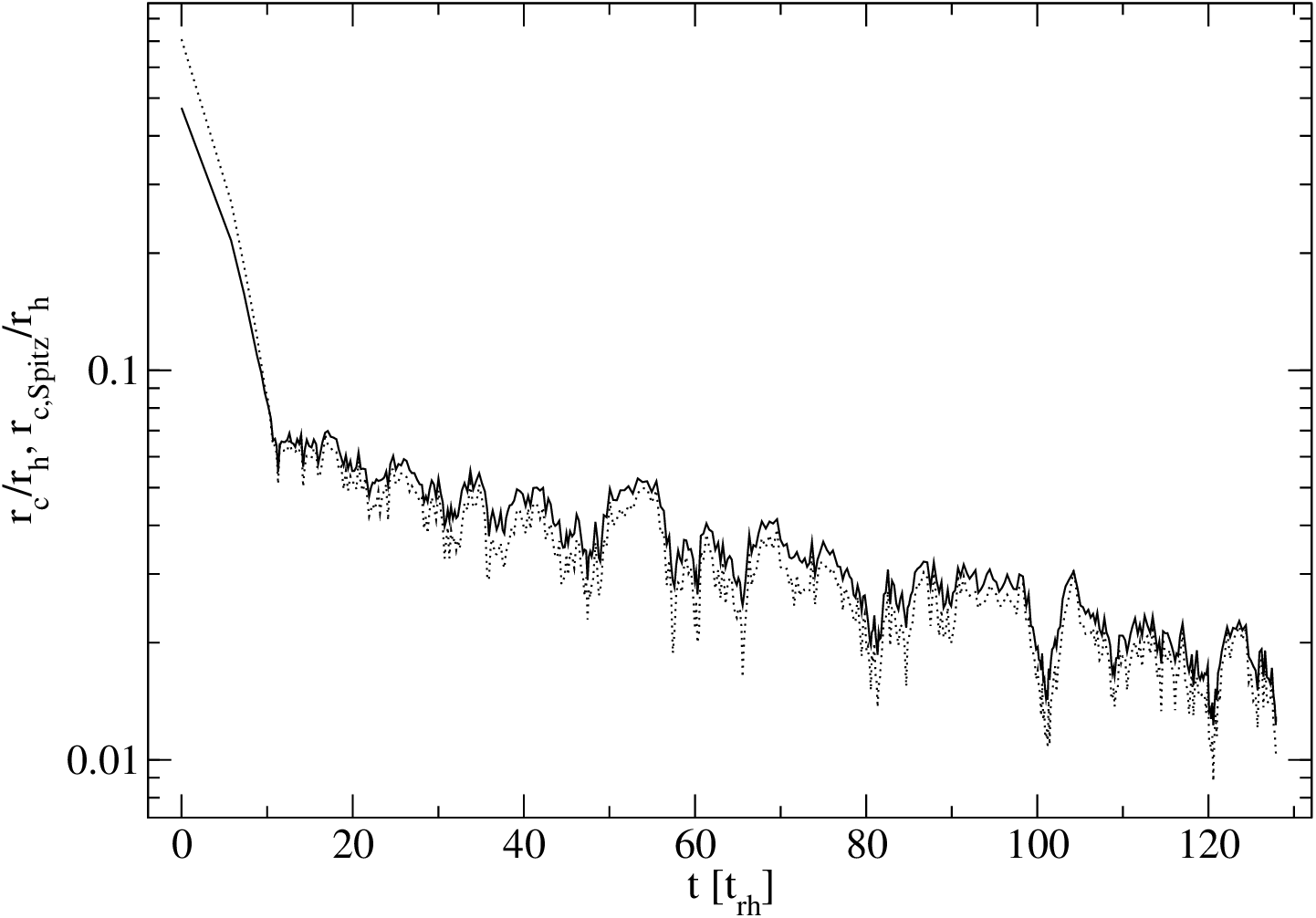}
    \caption{Evolution of $r_c/r_h$ for model T4.  The solid line uses the \citet{1985ApJ...298...80C}
      technique to measure the core radius, while the dotted line uses the standard \citet{1987degc.book.....S}
      definition.\label{fig:gao.rcrh}}
  \end{center}
\end{figure}

A standard model has emerged in the business of simulating the evolution of clusters 
containing primordial binaries.  The ``Gao, et al.''\ model is an isolated Plummer 
model with equal-mass stars
and 10\% binaries, with the binary binding energy distributed uniformly in the logarithm
from $3$ to $400kT$, and with the binary eccentricity distributed according to the
thermal distribution.  Since this model has been treated by all the previous work
cited above, we use it as a basis for comparison.  Fig.~\ref{fig:gao.binary} shows 
the evolution of this model (T4 in Table~\ref{tab:models}).  Fig.~\ref{fig:gao.rcrh}
shows the evolution of the ratio $r_c/r_h$.  Although we did not integrate our model
to deep core collapse, which occurs at $t \approx 50 t_{\rm rh}$ in \citet{1991ApJ...370..567G}
and $t \approx 130 t_{\rm rh}$ in \citet{2006MNRAS.368..677H}, 
there are still ample results for comparison.  Comparing with Fig.~1 of \citet{1991ApJ...370..567G},
the timescale of the initial core contraction we find is nearly identical
at $\approx 11 t_{\rm rh}$.  This is the same timescale found by \citet{2006MNRAS.368..677H}
in their Fig.~2 and \citet{2003MNRAS.343..781G} in their Fig.~4, 
although in the latter case the comparison is less meaningful
since their results show no long-lived binary burning phase.  It is also similar to the 
result of Paper III in Fig.~4, although in that paper we found a slightly 
shorter timescale.

Turning now to the size of the core radius during the binary burning phase, we find 
$r_c \approx 0.07$ in $N$-body units.  The model of \citet{1991ApJ...370..567G} shows
a smaller core size, with $r_c \approx 0.05$ in $N$-body units (note that the unit 
of length in their plot is $3\pi/16$ in $N$-body units).  This is somewhat expected, since the code of 
\citet{1991ApJ...370..567G} predicts a smaller core than the semi-analytical 
theory \citep{2006MNRAS.368..677H}.  The recipes-based model of Paper III shows a much larger core,
with $r_c \approx 0.2$.  This is not surprising, since recipes tend to overestimate the 
energy generation rate in binary interactions 
\citep{2003ApJ...593..772F,2003MNRAS.343..781G,astro-ph/0512032}.  Although the model
of \citet{2003MNRAS.343..781G} does not show a long-lived binary burning phase,
there is a subtle hint of a short-lived one with $r_c \approx 0.2$.  This is also 
much larger than the value we find.  Comparing our Fig.~\ref{fig:gao.rcrh} with
Fig.~2 of \citet{2006MNRAS.368..677H}, we see qualitatively very similar behavior
in the two calculations, 
with $r_c/r_h$ abruptly slowing its contraction at $t \approx 11 t_{\rm rh}$
and steadily decreasing thereafter.  At the start of the binary-burning phase,
\citet{2006MNRAS.368..677H} find $r_c/r_h \approx 0.06$, just as we do.
If the scaling of $r_c/r_h$ with $N$ in \citet{2006MNRAS.368..677H} can be
extrapolated out to $N \sim 10^5$ as described in section~\ref{sec:theorycomp},
then this agreement is somewhat surprising.  
In terms of the structural parameters during the binary burning phase
(which, as the longest-lived evolutionary phase in the life of a cluster is the most 
observationally relevant), as well as the timescale to reach it, our code appears to 
agree well with $N$-body, slightly less well with the Fokker-Planck
code, and much less well with the two other approximate codes.

We can also compare the mass in binaries (or, equivalently, number) as a function 
of time.  At the somewhat arbitrary time of $40 t_{\rm rh}$, we find 
$M_b/M_{b,0} \approx 0.35$.  \citet{1991ApJ...370..567G} find 
a value of 0.41, while \citet{2006MNRAS.368..677H} find
a value of 0.36.  In Paper III we found a value of 0.35.
All these methods appear to agree very well on the average rate that binaries
are lost (either ejected or disrupted in binary interactions) up to
$40 t_{\rm rh}$, although the agreement with Paper III is
probably fortuitous given the disagreement in the structural parameters.

\begin{figure}
  \begin{center}
    \includegraphics[width=\columnwidth]{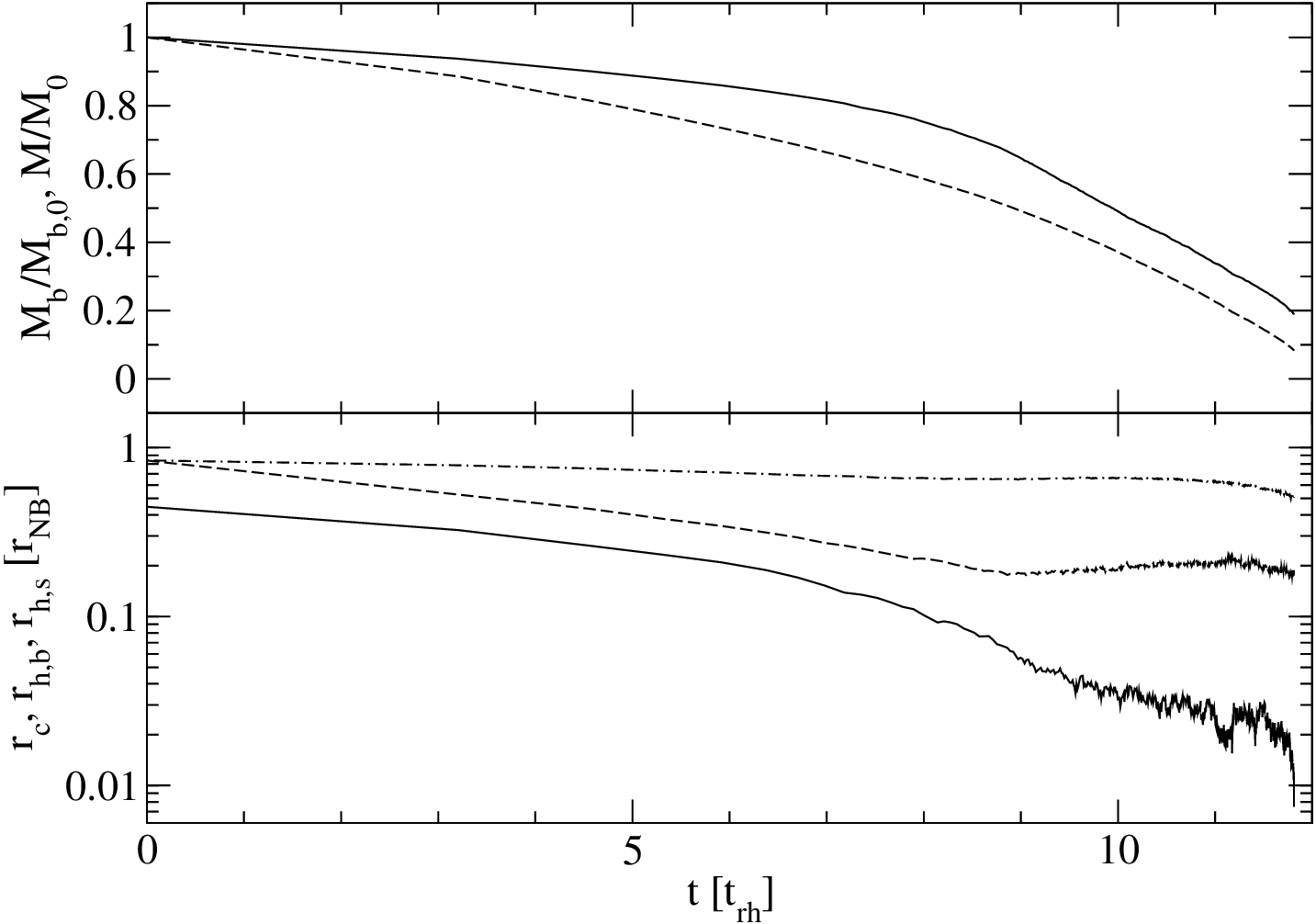}
    \caption{Evolution of model THH3.
      Quantities plotted are the same as in Fig.~\ref{fig:pl_n1e5_fb0.03.binary}.
      \label{fig:w3_n1e5_fb0.1.binary}}
  \end{center}
\end{figure}

\begin{figure}
  \begin{center}
    \includegraphics[width=\columnwidth]{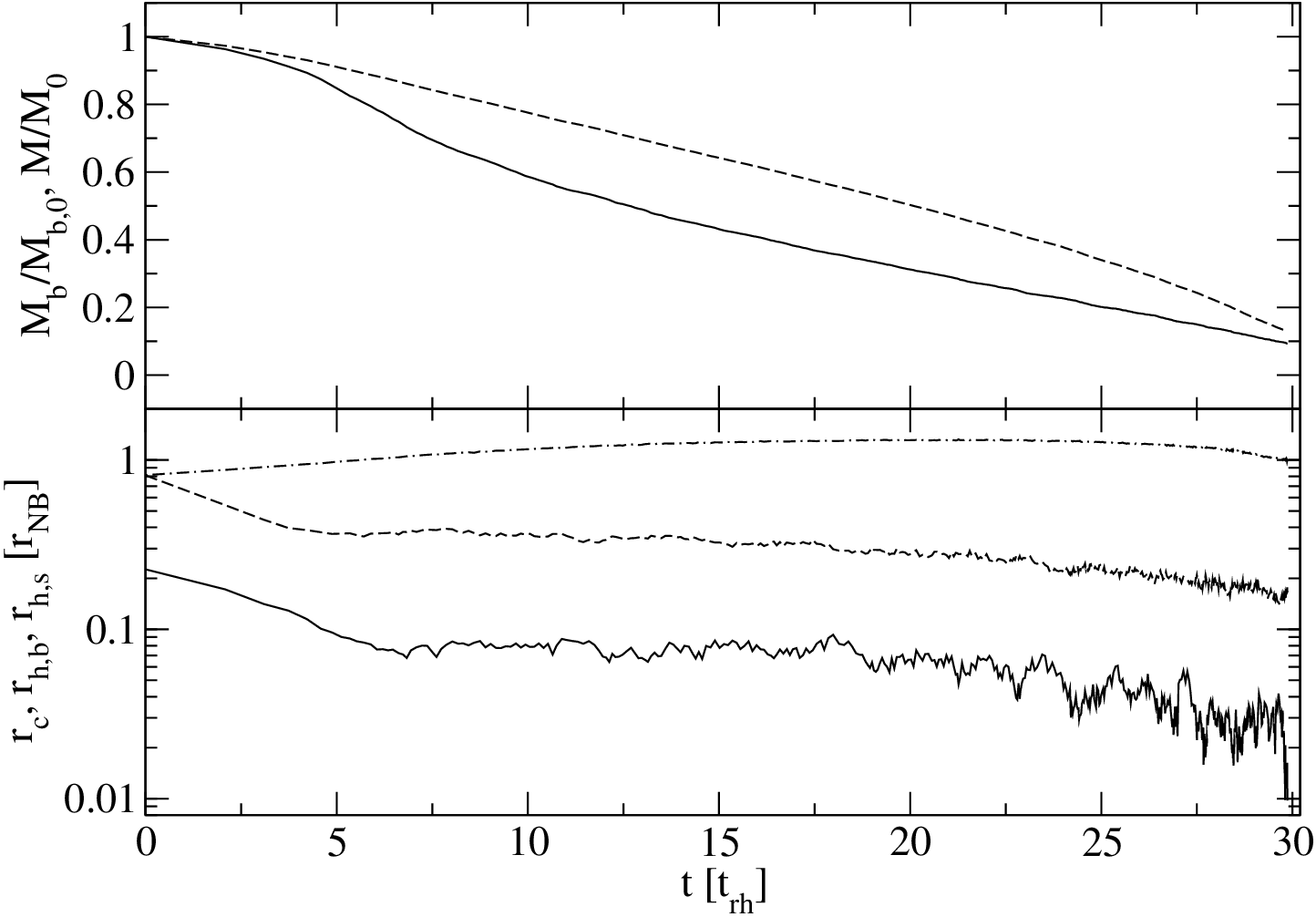}
    \caption{Evolution of model THH7.
      Quantities plotted are the same as in Fig.~\ref{fig:pl_n1e5_fb0.03.binary}.
      \label{fig:w7_n1e5_fb0.1.binary}}
  \end{center}
\end{figure}

Comparing an isolated cluster model tests in combination how 
well our code treats two-body relaxation, mass segregation, and 
the various aspects of binary burning.  Having shown that our code agrees
well with the ``exact'' method of $N$-body, one can be fairly confident that
our code treats these processes relatively accurately.  However, since all dense
star clusters are tidally truncated to some degree, it is still useful to consider
how well our code treats the physics of tidal stripping.  As described
in \citet{2001ApJ...550..691J}, we use a cutoff radius criterion to determine
whether a star has been stripped from the cluster.  This is in contrast to the
more accurate technique of including the tidal field in the equations of 
motion (which is not possible for the Monte Carlo method).  
\citet{astro-ph/0602409} have compared the two tidal stripping methods
by performing $N$-body simulations of $N \sim 10^4$ clusters with primordial binaries, 
and have found that models using a tidal cutoff tend to survive longer before 
disrupting (up to a factor of $\sim 2$, although it's not clear how this factor 
scales with $N$), and have a larger core radius (again, up to a factor of $\sim 2$),
than models with a tidal field.  We will keep this in mind when presenting results
below.  For now we compare two current simulations with the tidal cutoff models of 
\citet{astro-ph/0602409}, and with the previous version of our code.
Figs.~\ref{fig:w3_n1e5_fb0.1.binary} and \ref{fig:w7_n1e5_fb0.1.binary} show the 
evolution of models w3\_n1e5\_fb0.1 and w7\_n1e5\_fb0.1, respectively.  
Comparing the first with Fig.~13 in Paper III, we see that although both
models reach disruption, our new results are qualitatively different in behavior, yielding 
what appears to be a binary burning
phase from $9 t_{\rm rh}$ to disruption, while the old result shows no such phase.
Aside from this difference, the quantitative evolution of the structural
radii, the total cluster and binary mass, and the disruption timescale
appear to be very similar between the two models.
Comparing with Fig.~16 of \citet{astro-ph/0602409}, we find similar qualitative
behavior, with a hint of a binary burning phase in their results starting at roughly
the same time, similar evolution of the structural radii, and both models resulting
in disruption at $\sim 12$--$14 t_{\rm rh}$.  Our model seems to predict a smaller core radius,
by a factor of $\sim 2$.  Note that \citet{astro-ph/0602409} use the Spitzer definition 
of the core radius, which they find with their code to yield a value 
$\sim 20\%$ larger than the density-weighted averaging method.

Moving on to the $W_0=7$ model, we find similar qualitative behavior but shorter disruption 
times than both Fig.~10 in Paper III and \citet{astro-ph/0602409}
in their Fig.~17.  The recipes-based model of Paper III
predicts a much larger core radius in the binary burning phase than our current model.
This is due to the fact that the recipes used in that calculation tend to overestimate
the rate of energy generation in binary burning.  The agreement with \citet{astro-ph/0602409}
is better, with our core radius being only $\sim 20\%$ smaller than the $N$-body result.  Looking at
the evolution of $M$ and $M_b$, our model appears to exhibit very similar behavior to the 
$N$-body model.

In general, our new code yields much improved agreement with the 
$N$-body results reported in the literature, for both isolated and tidally-truncated
models, but yields some results that differ significantly from other approximate methods.  
Notably, the direct integration of binary interactions reduces their energy generation rate
relative to the simple recipes used in Paper III, and yields smaller core
radii.  This was evident to some degree in Paper III, in which the discrepancy
was illustrated for binary--single interactions.  Below we compare our new results
with observations and discuss the implications.

\section{Results and Comparison with Observations}\label{sec:results}

\begin{figure}
  \begin{center}
    \includegraphics[width=\columnwidth]{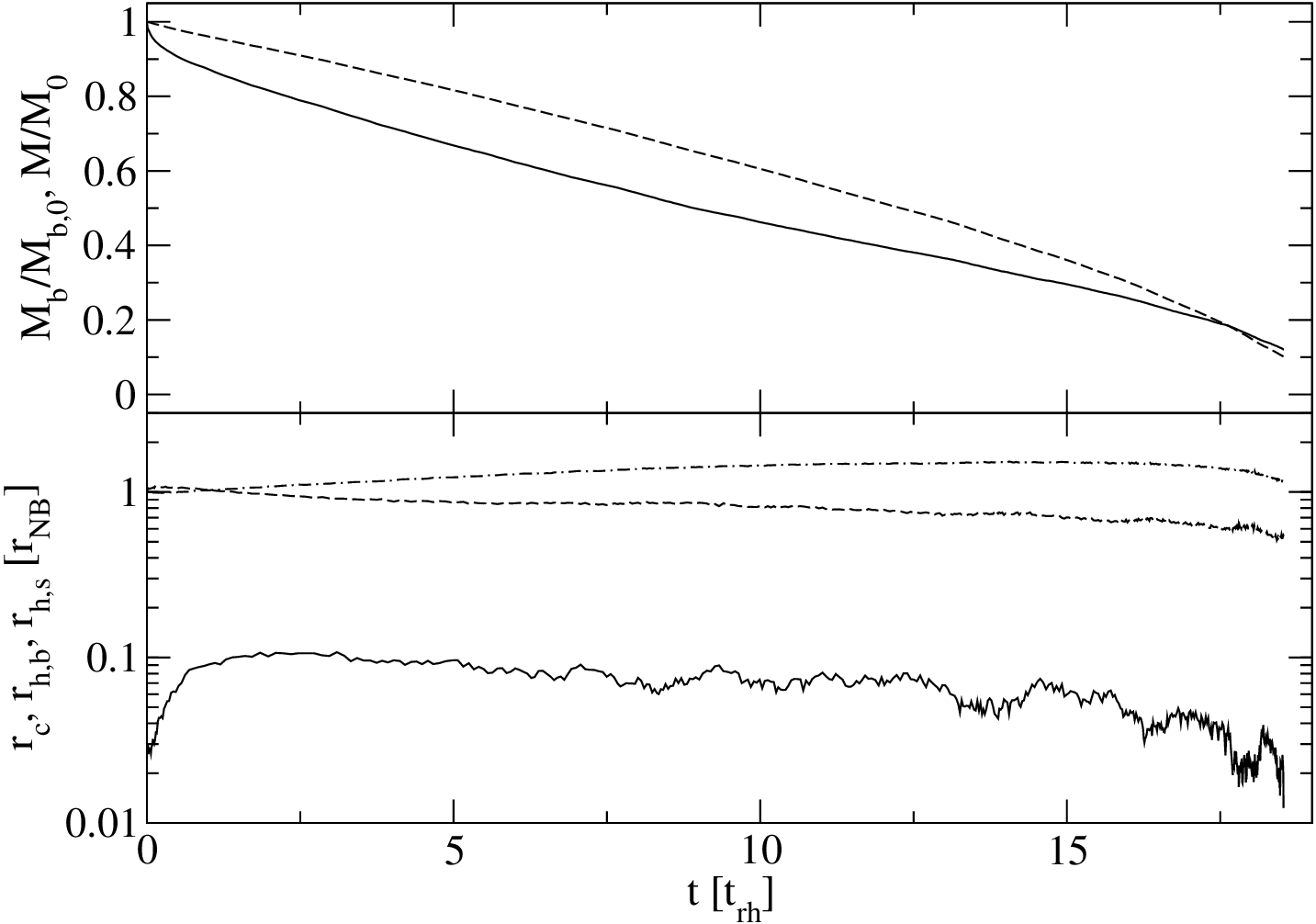}
    \caption{Evolution of model w11\_n1e5\_fb0.3.
      Quantities plotted are the same as in Fig.~\ref{fig:pl_n1e5_fb0.03.binary}.
      \label{fig:w11_n1e5_fb0.3.binary}}
  \end{center}
\end{figure}

\begin{figure}
  \begin{center}
    \includegraphics[width=\columnwidth]{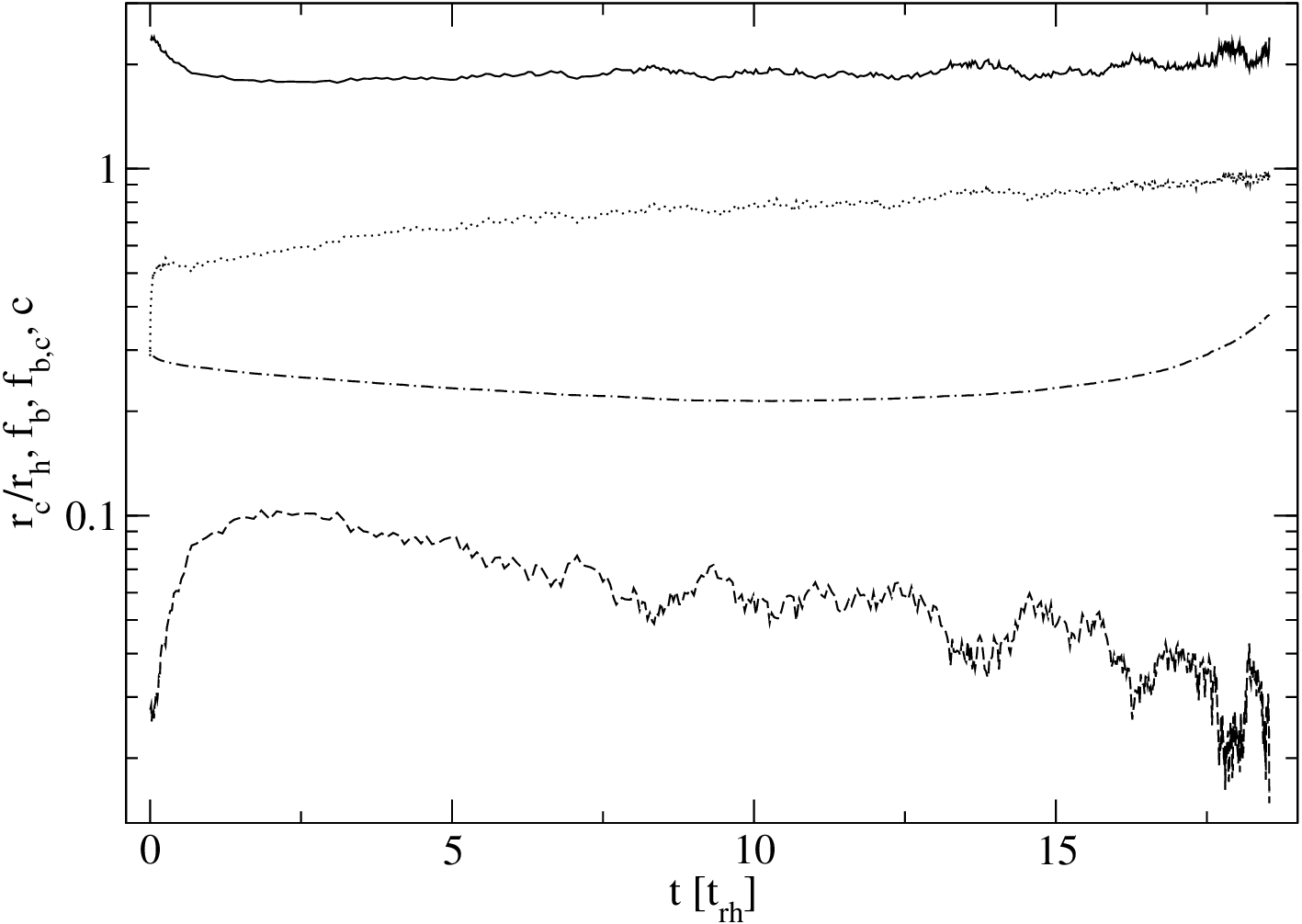}
    \caption{Evolution of $r_c/r_h$ (dashed line), total cluster binary fraction
      $f_b$ (dot-dashed line), core binary fraction $f_{\rm b,c}$ (dotted line), 
      and concentration $c$ (solid line) for model w11\_n1e5\_fb0.3.
      \label{fig:w11_n1e5_fb0.3.extras}}
  \end{center}
\end{figure}

\begin{figure}
  \begin{center}
    \includegraphics[width=\columnwidth]{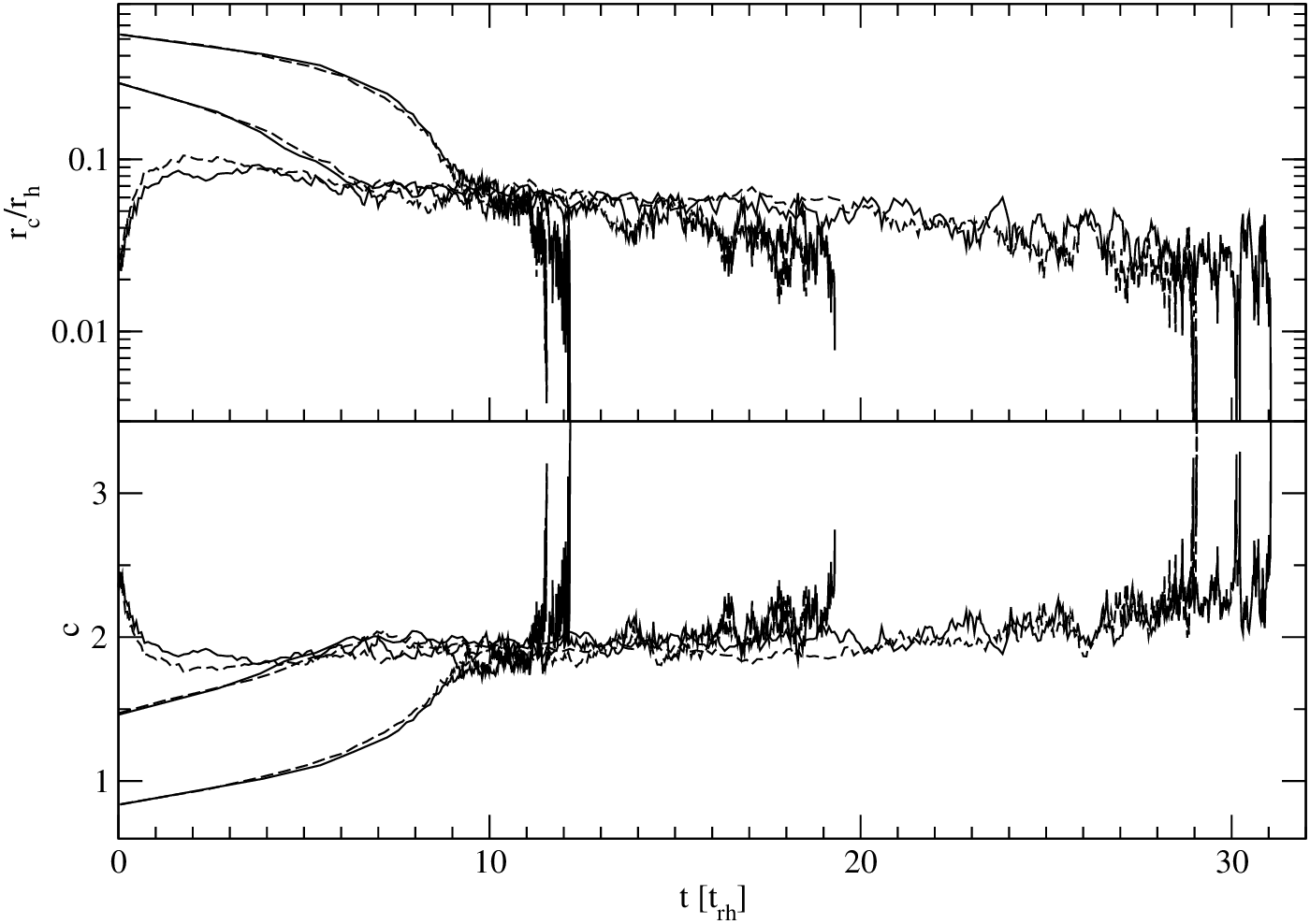}
    \caption{Evolution of $c$ and $r_c/r_h$ for $W_0=3$, $7$, and $11$ models
      with $f_b=0.1$ (solid lines) and $f_b=0.3$ (dashed lines).  Note the ``universal''
      behavior, with the initially more centrally concentrated models ($W_0=11$) expanding
      and the less centrally concentrated models ($W_0=3$ and $7$) contracting to common
      values of $r_c/r_h$ and $c$.\label{fig:r_rcrh_universal}}
  \end{center}
\end{figure}

\begin{figure}
  \begin{center}
    \includegraphics[width=\columnwidth]{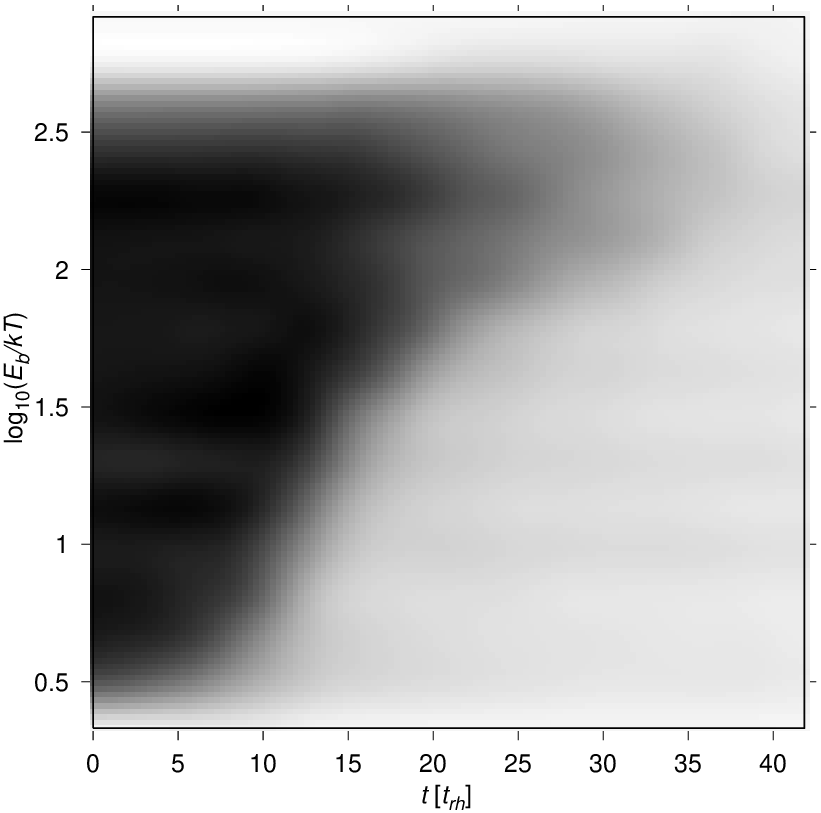}
    \caption{Evolution of the binary binding energy distribution for model T5.  The distribution starts
      off uniform in $\log E_b$ from 3 to 400 $kT$.  As the cluster evolves the central
      velocity dispersion increases, moving the hard--soft boundary to larger $E_b$ and destroying
      the wider binaries.  Near core collapse ($\sim 40 t_{\rm rh}$), only a small collection
      of relatively tight binaries ($E_b \sim 300\,kT$) remains.\label{fig:ebevol}}
  \end{center}
\end{figure}

In addition to displaying the initial conditions for all models simulated for this
paper, Table~\ref{tab:models} gives several important measured quantities
for each simulation.  The first is the core stabilization time, $t_{\rm cs}$, 
at which the core radius stabilizes (i.e., the start of the binary-burning 
phase) after the initial contraction or expansion.  Note that some authors denote this
as the core collapse time when discussing clusters with primordial binaries.  The second
is the time of the first deep core collapse, $t_{\rm cc}$, which, for models with
binaries, represents the time at which the binary population is nearly depleted in
the core.  The next is the disruption time, $t_{\rm dis}$, for models which are
tidally truncated.  The next is $r_c/r_h$, the ratio of the core to half-mass 
radius averaged over $1\,t_{\rm rh}$ after $t_{\rm cs}$.  Finally, 
$c=\log_{10}(r_t/r_c)$ is the concentration parameter averaged over the same time period.
The evolution of several of the models in the table has been shown graphically in the
preceding figures.  The behavior of the remaining models is similar to the ones already
shown, with the relevant timescales and structural parameters appropriately modified.  The
only exceptions are the high concentration King models, which undergo an initial phase
of core expansion (instead of contraction) due to their initially very dense state.  
Fig.~\ref{fig:w11_n1e5_fb0.3.binary} displays the evolution of such a model, model
w11\_n1e5\_fb0.3.  For reference we have plotted the time evolution of the measured
quantities $r_c/r_h$ and $c$ in Fig.~\ref{fig:w11_n1e5_fb0.3.extras}, along with the 
the total cluster binary fraction $f_b$ and the core binary fraction $f_{\rm b,c}$.

Fig.~\ref{fig:r_rcrh_universal} shows the evolution of $c$ and $r_c/r_h$ for $W_0=3$, $7$, and $11$ models
with $f_b=0.1$ (solid lines) and $f_b=0.3$ (dashed lines).  Of note is the ``universal'' behavior
displayed by the models, in which the initially more centrally concentrated models ($W_0=11$) expand
and the less centrally concentrated models ($W_0=3$ and $7$) contract to common
values of $r_c/r_h$ and $c$.  This is the same behavior found previously in the literature,
in Paper III and \citet{astro-ph/0602409} in their Fig.~8, although in the latter 
case they find systematically larger values of $r_c/r_h$ and smaller values of $c$ (as described above).
For reference we also show in Fig.~\ref{fig:ebevol} the evolution of the binary binding energy 
distribution for model T5.  The softest binaries begin to be destroyed (by ejection or disruption
in binary interactions) on the approach to core stabilization.  As the cluster evolves the central
velocity dispersion increases, moving the hard--soft boundary to larger $E_b$ and destroying
the wider binaries.  Near core collapse ($\sim 40 t_{\rm rh}$), only a small collection
of relatively tight binaries ($E_b \sim 300\,kT$) remains.  The evolution of the binary population
in $E_b$--$r$ space is qualitatively similar to what's shown in Fig.~15 of Paper III, with
the softer binaries first being destroyed in the core, and the harder binaries becoming
even harder and more centrally concentrated with time.

There are several trends apparent in the data presented in Table~\ref{tab:models}.  
For fixed initial cluster structure,
as $f_b$ is increased from zero, $t_{\rm cs}$ decreases from the value of $t_{\rm cc}$ at $f_b=0$,
reaches a minimum around $f_b \sim 0.2$, then increases back to approximately its $f_b=0$ value
for $f_b=1$.  This dip is due to the fact that in the initial core contraction or expansion phase, 
binaries act primarily as a second, heavier star species, hastening cluster energy transport
via mass segregation.  Looking at the deep core collapse times, another clear trend
is that $t_{\rm cc}$ increases dramatically as $f_b$ is increased.  This trend is most
striking in the isolated Plummer models, with merely $f_b \approx 0.03$ being enough to 
double the deep core collapse time, and $f_b=1$ increasing it
by a factor of at least 7 (model T4).  The physical explanation is, of course, that the
lifetime of the binary-burning phase increases with the amount of fuel available,
in analogy to hydrogen-burning in main-sequence stars.  Looking more carefully
at the tidally-truncated King models, we see that the presence of binaries tends to drive
these clusters to complete tidal disruption.  The minimum of $t_{\rm dis}$ 
occurs somewhere in the range $0.1 \lesssim f_b \lesssim 1$.  This is not surprising, since
the maximum of $r_c/r_h$ occurs at $f_b \approx 0.4$ (see Fig.~\ref{fig:vesp_chern}),
implying that the cluster is the most distended for this value of $f_b$.  Moving
now to the cluster structural parameters, we see that there is relatively little variation
in $r_c/r_h$ and $c$ over the wide range of cluster initial profiles and binary fractions considered, with
$r_c/r_h$ ranging from $\sim 0.05$ for Plummer or $W_0 \sim 7$ King models with low $f_b$,
to $\sim 0.1$ for larger $W_0$ King models with $f_b \sim 1$.  The concentration
parameter shows a similarly small amount of variation, peaking at $\sim 2.1$ for $W_0 \sim 7$ and
low binary fraction, and falling to $\sim 1.7$ for larger $W_0$ and larger
binary fraction.  As expected, a larger $f_b$ leads to a smaller $c$, since the core radius 
generally increases with binary fraction.  Another trend is evident when comparing single-mass
models with models incorporating a more realistic mass spectrum (Salpeter from $0.2$--$1.2\,M_\sun$):
models with a mass spectrum tend to have a slightly smaller $r_c/r_h$, and show more
variation of $c$ with $W_0$.

Although our cluster evolution models are rather simplified (since they do not include single-
or binary-star stellar evolution, or collisions), it is still useful to compare the predicted structural
parameters with observations.  In Paper III we compared $r_c/r_h$ and $c$
from simulations using the previous version of our Monte Carlo code (which includes
recipes for binary interactions instead of dynamical integrations), with observations for the
Galactic globular clusters.  There we found promising agreement, with $r_c/r_h$ from the
simulations falling generally in the low $r_c/r_h$ region of the observed distribution for non-core collapsed
clusters, which extends from $\sim 0.1$ to $\sim 1$ with a peak at $\sim 0.5$.  And similarly for $c$, 
falling in the high $c$ region of the observed distribution for non-core collapsed clusters, which extends from 
$\sim 0.5$ to $\sim 2.5$ with a peak at $\sim 1.5$.  However, as described above, we find
systematically smaller values for $r_c/r_h$ than in Paper III by a factor of up to $\sim 10$,
as well as systematically larger values for $c$ by $\sim 0.5$.  
The improved treatment of binary interactions in our code has shifted our
predictions for $r_c/r_h$ and $c$ outside the observed ranges for non-core collapsed clusters,
now yielding agreement only with the roughly 10\% of Galactic globular clusters that are 
classified observationally as core collapsed.

A globular cluster of comparable mass stars with $f_b \gtrsim 0.03$ viewed at a random
time during its life has a greater than 50\% chance of being found in the quasi-steady state
binary burning phase.  When one considers that globulars are likely born with significant
binary fractions \citep{1992PASP..104..981H,2005MNRAS.358..572I}, and that we are currently
observing the Galactic globular clusters at a late stage in their evolution, 
the vast majority of observed clusters should currently be in the binary-burning phase.  
The most obvious interpretation of the observational data is that most Galactic globular clusters
are currently in the binary-burning phase, and the roughly 10\% classified
as core-collapsed are within a small time window around a
deep core collapse phase.  The disagreement between simulations and observations for the 
structural parameters of the non core-collapsed clusters, then, suggests one of at least two possibilities: 1)
the Galactic globular clusters do not start within the volume in parameter space of initial conditions
we have considered here; or 2) there are additional physical processes at work in clusters,
yielding larger cores.  If clusters are born with $W_0 \gtrsim 11$ and $f_b \lesssim 1$, 
extrapolation of our results suggests that simulations may then agree with the observations of
non-core collapsed clusters, implying possibility (1) may be correct.  
However, a cluster with $W_0 \gtrsim 11$ is likely to have a short enough central
relaxation time that a runaway stellar collision will occur, creating an intermediate
mass black hole (IMBH) early in the cluster's lifetime \citep{2006MNRAS.368..141F}.  
This is intriguing, since clusters with central IMBHs typically have significantly larger
values of $r_c/r_h$ than clusters without \citep{astro-ph/0610342}.
We have not included in our simulations any form of stellar evolution (for single stars or binaries),
or physical stellar collisions.  Single star evolution tends to heat a cluster early in its lifetime
via wind-driven mass loss and supernovae explosions, causing the cluster and its core to expand.
However, this effect is most pronounced only early in the lifetime of a star cluster 
($\lesssim 1\,{\rm Gyr}$).  The effects of binary stellar evolution are less obvious, since
it is a rather complicated process.  However, the simulations of \citet{2005MNRAS.358..572I}
suggest that the binary fraction in the core quickly drops to relatively small
values ($\lesssim 20\%$), and that binary stellar evolution tends to destroy tight binaries.
The net result is likely to be a smaller equilibrium value of $r_c/r_h$ \citep{1994ApJ...431..231V} 
than with no binary stellar evolution, suggesting that possibility (2) is less likely. 
A refined quantitative study combining the results
of \citet{2005MNRAS.358..572I} and \citet{1994ApJ...431..231V} would more clearly elucidate this
effect.  

The effect of direct single--single star collisions in young dense clusters is to dissipate orbital
energy and drive core collapse \citep{2006MNRAS.368..141F}.  For clusters in which stellar
merger products have had time to evolve and lose mass through accelerated stellar evolution,
the net result may be to heat the core \citep{1987ApJ...319..801L}.  The degree to which this
process operates in Galactic globular clusters is unclear, however.
The effect of stellar collisions during binary interactions is generally to reduce the
efficiency of binary burning \citep{1985ApJ...298..502H,1986ApJ...306..552M,1991ApJ...378..637G}.
This is because a merger product resulting from
a star--star collision typically has significantly more internal energy (potential and rotational)
than the sum of the merging stars' internal energies.  Thus when a collision occurs during a 
binary dynamical interaction, effectively some of the binding energy of the binary is converted
into stellar binding energy, decreasing the efficiency of binary burning.
In other words, energy that had previously been available to be
converted into kinetic energy through binary interactions is no longer available, tied up 
in the stars.  The result should be a decreased value of $r_c/r_h$ in the binary-burning phase relative to
the case of point-mass binary dynamics, implying that
possibility (2) is less likely.  As with binary stellar evolution, however, a more detailed simulation
which studies the effects of collisions in an evolving model should be performed to quantify the effect.

\section{Summary and Conclusions}\label{sec:summary}

In this paper we have described our new Monte Carlo evolution code and used it
to perform a large set of cluster evolution simulations, which we compared 
with previous results in the literature, as well as observations of Galactic 
globular clusters.

In section \ref{sec:method} we described our new code in detail, including
the implementation of direct integration of binary scattering interactions
and star--star physical collisions, and the fundamental modifications we have made to the 
core Monte Carlo method.  We performed several test calculations
with the code, finding that it reproduces well several standard results.  It yields
a core collapse time of $\approx 18\,t_{\rm rh}$ for an isolated Plummer model,
in good agreement with results in the literature.  It also produces an $r^{-2.3}$
density profile during the late stages of core collapse, in good agreement with the 
theoretical expectation.  We also performed comparisons of clusters with increasingly
wide mass spectra with $N$-body, finding that for moderately wide mass
spectra ($1$ to $10\,M_\sun$) the agreement with $N$-body is satisfactory, but
for very wide mass spectra ($0.2$ to $120\,M_\sun$) the agreement is not as good.
In particular, for such wide mass spectra, our Monte Carlo code tends to overestimate
the mass segregation timescale at early times, and underestimate it at later times.
The sense of the disagreement is the same as found by \citet{2006MNRAS.368..121F} with their
Monte Carlo code.

In section \ref{sec:exampleresults} we displayed a few example results and compared
with theory and previous numerical calculations in the literature.  We found that the code
conserves energy well over the long timescales of our runs.  
We compared our predicted values of $r_c/r_h$ during the quasi-steady
state binary burning phase with the semi-analytical work of \citet{1994ApJ...431..231V},
finding good agreement.  We compared our results with previous
numerical results in the literature for isolated and tidally-truncated cluster models,
finding excellent agreement with $N$-body calculations.  There are much larger
discrepancies with the other approximate methods (Fokker-Planck and other Monte Carlo codes),
which is to be expected, since most used recipes for binary interactions, which are known
to overestimate the energy generation rate in binaries.

In section \ref{sec:results} we surveyed the results from all our simulations, and compared
with observations.  Our models cover a large range in parameter space, using Plummer and King models
with $W_0=3$ to 11 for the initial cluster profile, and with initial binary fractions from
0 to 1.  The resulting structural parameters in the binary burning phase span a remarkably small
range, with $r_c/r_h$ varying from 0.03 to 0.12, and $c$ varying from 1.7 to 2.4.
Our results for these structural parameters are distinctly different from the results found
with the previous version of our code (which used recipes for binary interactions), with 
$r_c/r_h$ now smaller than what we found in Paper III by a factor of up to $\sim 10$,
and with $c$ larger by $\sim 0.5$.  Although our new results agree much better with
$N$-body calculations, they unfortunately agree much less well than in 
Paper III with the observations.  The disagreement implies one
of at least two possibilities.  It may be that the initial conditions for Galactic globular 
clusters are outside the range of initial conditions we have sampled in this work.   Extrapolation
of our results suggests that clusters with $W_0 \gtrsim 11$ and $f_b \lesssim 1$
may match the observations.  This is intriguing since a cluster with $W_0 \gtrsim 11$
will likely form an IMBH early in its lifetime via a collisional runaway, and clusters
with central IMBHs are known to have larger values of $r_c/r_h$ \citep{2006MNRAS.368..141F,astro-ph/0610342}.
Alternatively, stellar evolution and collisions, which are
not included in the simulations in this paper, could possibly explain the disagreement.
However, it appears that the effect of these processes should act in the opposite sense
of ameliorating the disagreement with observations.  More detailed simulations including
the effects of single- and binary-star evolution and physical collisions should be
performed to test this.

\acknowledgements

We thank the referee, Piet Hut, for many comments and suggestions that greatly improved
this work.  We also thank Michele Trenti for providing the code to calculate $r_c/r_h(\phi_b)$
in \citet{1994ApJ...431..231V}, and Holger Baumgardt for providing
the $N$-body data for Fig.~\ref{fig:plummer_lagrad} and 
Figs.~\ref{fig:plummer_kroupa_lagrad} through \ref{fig:plummer_salpeter_avemass}.
For illuminating discussions and comments on the manuscript we thank 
Marc Freitag, M. Atakan G\"urkan, Douglas Heggie, Craig Heinke, and Michele Trenti.
The authors acknowledge support from NASA Grant NNG06GI62G.

\bibliographystyle{apj}
\bibliography{apj-jour,main}

\clearpage

\begin{deluxetable}{crccclcccccc}
  \tablecaption{Parameters and measured quantities for all model simulations presented in this paper.\label{tab:models}}
  \tablehead{
    \colhead{name} &
    \colhead{$N$} &
    \colhead{profile} &
    \colhead{$r_{\rm NB}/{\rm pc}$} &
    \colhead{$f(M/M_\sun)$} &
    \colhead{$f_b$} &
    \colhead{$f(E_b/kT)$} &
    \colhead{$t_{\rm cs}/t_{\rm rh}$} &
    \colhead{$t_{\rm cc}/t_{\rm rh}$} &
    \colhead{$t_{\rm dis}/t_{\rm rh}$} &
    \colhead{$r_c/r_h$} &
    \colhead{$c$}
  }
  \startdata
  T1                   &$5 \times 10^5$    &Plum.  &$1.02$ &$\propto \delta(M-1)$      &0    &\nodata                         &\nodata &17.6 &\nodata &\nodata &\nodata\\
  T2                   &$10^6$             &Plum.  &$0.58$ &Kroup., $0.1$--$10$        &0    &\nodata                         &\nodata &0.54 &\nodata &\nodata &\nodata\\
  T3                   &$1.25 \times 10^6$ &Plum.  &$0.60$ &Salp., $0.2$--$120$        &0    &\nodata                         &\nodata &0.067 &\nodata &\nodata &\nodata\\
  T4                   &$10^5$             &Plum.  &$7.49$ &$\propto \delta(M-1)$      &0.1  &$\propto E_b^{-1}$, $3$--$400$  &11      &$>128$ &\nodata &0.06 &\nodata\\
  T5                   &$10^5$             &Plum.  &$7.33$ &$\propto \delta(M-1)$      &0.03  &$\propto E_b^{-1}$, $3$--$400$  &14      &$42$ &\nodata &0.06 &\nodata\\
  \hline
  THH3                 &$10^5$             &$W_0=3$ &$7.49$ &$\propto \delta(M-1)$     &0.1  &$\propto E_b^{-1}$, $3$--$400$  &$9.4$ &$11.8$ &$>11.8$ &$0.07$ &$1.8$\\
  THH7                 &$10^5$             &$W_0=7$ &$7.49$ &$\propto \delta(M-1)$     &0.1  &$\propto E_b^{-1}$, $3$--$400$  &$6.7$ &$29.8$ &$>29.8$ &$0.08$ &$1.9$\\
  \hline
  pl\_n3e5\_fb0.01\_kt &$3 \times 10^5$    &Plum.  &$5.44$ &$\propto \delta(M-1)$      &0.01 &$\propto E_b^{-1}$, $10$--$100$ &$16.3$ &$>23$ &\nodata &$0.04$ &\nodata\\
  pl\_n3e5\_fb0.02\_kt &$3 \times 10^5$    &Plum.  &$5.46$ &$\propto \delta(M-1)$      &0.02 &$\propto E_b^{-1}$, $10$--$100$ &$14.2$ &$>23$ &\nodata &$0.06$ &\nodata\\
  pl\_n3e5\_fb0.04\_kt &$3 \times 10^5$    &Plum.  &$5.49$ &$\propto \delta(M-1)$      &0.04 &$\propto E_b^{-1}$, $10$--$100$ &$12.5$ &$>23$ &\nodata &$0.07$ &\nodata\\
  pl\_n3e5\_fb0.08\_kt &$3 \times 10^5$    &Plum.  &$5.56$ &$\propto \delta(M-1)$      &0.08 &$\propto E_b^{-1}$, $10$--$100$ &$11.0$ &$>23$ &\nodata &$0.07$ &\nodata\\
  pl\_n3e5\_fb0.15\_kt &$3 \times 10^5$    &Plum.  &$5.68$ &$\propto \delta(M-1)$      &0.15 &$\propto E_b^{-1}$, $10$--$100$ &$10.4$ &$>23$ &\nodata &$0.06$ &\nodata\\
  pl\_n3e5\_fb0.30\_kt &$3 \times 10^5$    &Plum.  &$5.92$ &$\propto \delta(M-1)$      &0.3  &$\propto E_b^{-1}$, $10$--$100$ &$12.6$ &$>23$ &\nodata &$0.06$ &\nodata\\
  pl\_n3e5\_fb0.60\_kt &$3 \times 10^5$    &Plum.  &$6.34$ &$\propto \delta(M-1)$      &0.6  &$\propto E_b^{-1}$, $10$--$100$ &$17.5$ &$>23$ &\nodata &$0.05$ &\nodata\\
  \hline
  pl\_n1e5\_fb0        &$10^5$             &Plum.  &$7.25$ &$\propto \delta(M-1)$      &0    &\nodata                         &\nodata &$17.6$ &\nodata &\nodata &\nodata\\
  pl\_n1e5\_fb0.03     &$10^5$             &Plum.  &$7.33$ &$\propto \delta(M-1)$      &0.03 &$\propto E_b^{-1}$, phys.       &$15$ &$33.3$ &\nodata &$0.05$ &\nodata\\
  pl\_n1e5\_fb0.1      &$10^5$             &Plum.  &$7.49$ &$\propto \delta(M-1)$      &0.1  &$\propto E_b^{-1}$, phys.       &$10$ &$>28$ &\nodata &$0.07$ &\nodata\\
  pl\_n1e5\_fb0.3      &$10^5$             &Plum.  &$7.92$ &$\propto \delta(M-1)$      &0.3  &$\propto E_b^{-1}$, phys.       &$12$ &$>32$ &\nodata &$0.06$ &\nodata\\
  pl\_n1e5\_fb1        &$10^5$             &Plum.  &$9.14$ &$\propto \delta(M-1)$      &1    &$\propto E_b^{-1}$, phys.       &$18$ &$>22$ &\nodata &$0.07$ &\nodata\\
  \hline
  w3\_n1e5\_fb0        &$10^5$             &$W_0=3$  &$7.25$ &$\propto \delta(M-1)$      &0    &\nodata                       &\nodata &$13.7$ &$>13.7$ &\nodata &\nodata\\
  w3\_n1e5\_fb0.03     &$10^5$             &$W_0=3$  &$7.33$ &$\propto \delta(M-1)$      &0.03 &$\propto E_b^{-1}$, phys.     &$11$ &$13.8$ &$13.8$ &$0.06$ &$1.9$\\
  w3\_n1e5\_fb0.1      &$10^5$             &$W_0=3$  &$7.49$ &$\propto \delta(M-1)$      &0.1  &$\propto E_b^{-1}$, phys.     &$9$ &$12.2$ &$12.2$ &$0.08$ &$1.8$\\
  w3\_n1e5\_fb0.3      &$10^5$             &$W_0=3$  &$7.92$ &$\propto \delta(M-1)$      &0.3  &$\propto E_b^{-1}$, phys.     &$10$ &$11.5$ &$11.5$ &$0.06$ &$1.8$\\
  w3\_n1e5\_fb1        &$10^5$             &$W_0=3$  &$9.14$ &$\propto \delta(M-1)$      &1    &$\propto E_b^{-1}$, phys.     &$14$ &$16.3$ &$16.3$ &$0.06$ &$1.9$\\
  \hline
  w7\_n1e5\_fb0        &$10^5$             &$W_0=7$  &$7.25$ &$\propto \delta(M-1)$      &0    &\nodata                       &\nodata &$11.1$ &$>11.1$ &\nodata &\nodata\\
  w7\_n1e5\_fb0.03     &$10^5$             &$W_0=7$  &$7.33$ &$\propto \delta(M-1)$      &0.03 &$\propto E_b^{-1}$, phys.     &$10$ &$19.9$ &$>19.9$ &$0.05$ &$2.1$\\
  w7\_n1e5\_fb0.1      &$10^5$             &$W_0=7$  &$7.49$ &$\propto \delta(M-1)$      &0.1  &$\propto E_b^{-1}$, phys.     &$7$ &$30.3$ &$>30.3$ &$0.07$ &$2.0$\\
  w7\_n1e5\_fb0.3      &$10^5$             &$W_0=7$  &$7.92$ &$\propto \delta(M-1)$      &0.3  &$\propto E_b^{-1}$, phys.     &$7$ &$29.0$ &$>29.0$ &$0.06$ &$2.0$\\
  w7\_n1e5\_fb1        &$10^5$             &$W_0=7$  &$9.14$ &$\propto \delta(M-1)$      &1    &$\propto E_b^{-1}$, phys.     &$14$ &$38.7$ &$38.7$ &$0.07$ &$1.9$\\
  \hline
  w11\_n1e5\_fb0       &$10^5$             &$W_0=11$ &$7.25$ &$\propto \delta(M-1)$      &0    &\nodata                       &\nodata &$1.0$ &$>1.0$ &\nodata &\nodata\\
  w11\_n1e5\_fb0.03    &$10^5$             &$W_0=11$ &$7.33$ &$\propto \delta(M-1)$      &0.03 &$\propto E_b^{-1}$, phys.     &$2$ &$13.0$ &$>13.0$ &$0.07$ &$2.0$\\
  w11\_n1e5\_fb0.1     &$10^5$             &$W_0=11$ &$7.49$ &$\propto \delta(M-1)$      &0.1  &$\propto E_b^{-1}$, phys.     &$1$ &$>3.6$ &$>3.6$ &$0.09$ &$1.9$\\
  w11\_n1e5\_fb0.3     &$10^5$             &$W_0=11$ &$7.92$ &$\propto \delta(M-1)$      &0.3  &$\propto E_b^{-1}$, phys.     &$1$ &$>2.9$ &$>2.9$ &$0.09$ &$1.8$\\
  w11\_n1e5\_fb1       &$10^5$             &$W_0=11$ &$9.14$ &$\propto \delta(M-1)$      &1    &$\propto E_b^{-1}$, phys.     &$2$ &$25.0$ &$25.0$ &$0.12$ &$1.7$\\
  \hline
  pl\_n1e5\_s\_fb0     &$10^5$             &Plum.  &$5.05$ &Salp., $0.2$--$1.2$ &0    &\nodata                     &\nodata &4.9 &\nodata &\nodata &\nodata\\
  pl\_n1e5\_s\_fb0.03  &$10^5$             &Plum.  &$5.09$ &Salp., $0.2$--$1.2$ &0.03 &$\propto E_b^{-1}$, phys.   &$5$ &$12.7$ &\nodata &$0.03$ &\nodata\\
  pl\_n1e5\_s\_fb0.1   &$10^5$             &Plum.  &$5.17$ &Salp., $0.2$--$1.2$ &0.1  &$\propto E_b^{-1}$, phys.   &$5$ &$>26$ &\nodata &$0.05$ &\nodata\\
  pl\_n1e5\_s\_fb0.3   &$10^5$             &Plum.  &$5.40$ &Salp., $0.2$--$1.2$ &0.3  &$\propto E_b^{-1}$, phys.   &$4$ &$>67$ &\nodata &$0.09$ &\nodata\\
  \hline
  w4\_n1e5\_s\_fb0    &$10^5$             &$W_0=4$  &$5.05$ &Salp., $0.2$--$1.2$ &0    &\nodata                     &\nodata &5.4 &\nodata &\nodata &\nodata\\
  w4\_n1e5\_s\_fb0.03 &$10^5$             &$W_0=4$  &$5.09$ &Salp., $0.2$--$1.2$ &0.03 &$\propto E_b^{-1}$, phys.   &$5$ &$7.1$ &$>7.1$ &$0.04$ &$2.1$\\
  w4\_n1e5\_s\_fb0.1  &$10^5$             &$W_0=4$  &$5.17$ &Salp., $0.2$--$1.2$ &0.1  &$\propto E_b^{-1}$, phys.   &$5$ &$8.1$ &$8.1$ &$0.06$ &$1.9$\\
  w4\_n1e5\_s\_fb0.3  &$10^5$             &$W_0=4$  &$5.40$ &Salp., $0.2$--$1.2$ &0.3  &$\propto E_b^{-1}$, phys.   &$6$ &$8.6$ &$8.6$ &$0.06$ &$1.8$\\
  \hline
  w8\_n1e5\_s\_fb0    &$10^5$             &$W_0=8$  &$5.05$ &Salp., $0.2$--$1.2$ &0    &\nodata                     &\nodata &1.0 &\nodata &\nodata &\nodata\\
  w8\_n1e5\_s\_fb0.03 &$10^5$             &$W_0=8$  &$5.09$ &Salp., $0.2$--$1.2$ &0.03 &$\propto E_b^{-1}$, phys.   &$1$ &$7.2$ &$>7.2$ &$0.03$ &$2.4$\\
  w8\_n1e5\_s\_fb0.1  &$10^5$             &$W_0=8$  &$5.17$ &Salp., $0.2$--$1.2$ &0.1  &$\propto E_b^{-1}$, phys.   &$1$ &$>7.6$ &$>7.6$ &$0.06$ &$2.2$\\
  w8\_n1e5\_s\_fb0.3  &$10^5$             &$W_0=8$  &$5.40$ &Salp., $0.2$--$1.2$ &0.3  &$\propto E_b^{-1}$, phys.   &$1$ &$20.2$ &$20.2$ &$0.07$ &$2.1$\\
  \enddata
  \tablecomments{Here $N$ is the number of total cluster objects (single stars and binaries), 
    the profile is either a Plummer model or a King model with the specified $W_0$,
    $r_{\rm NB}$ is the unit of length in the simulation, $f(M)$ is the initial mass function,
    $f_b$ is the initial binary fraction, $f(E_b)$ is the distribution of binary binding energy, 
    $t_{\rm cs}$ is the time at which the core radius stabilizes (i.e., the start of the binary-burning 
    phase) after the initial contraction or expansion (note that some authors denote this
    as the core collapse time when discussing clusters with primordial binaries), $t_{\rm cc}$ is the time of the
    first deep core collapse, $t_{\rm dis}$ is the time at which the cluster disrupts due to
    tidal stripping, $r_c/r_h$ is the ratio of the core to half-mass radius averaged over 
    $1 t_{\rm rh}$ after $t_{\rm cs}$, and $c=\log_{10}(r_t/r_c)$ is the concentration parameter
    averaged over the same time period.  Quantities are omitted when the physical state they describe is 
    never reached (or can never be reached) during the simulation.  Note that those models 
    with $f_b=0$ or those with $kT$-based limits on $E_b$ have one degree of freedom in their
    scaling, while those with physical limits on the binary population cannot be rescaled.}
\end{deluxetable}

\end{document}